\documentclass[preprint,10pt]{revtex4-1}
\usepackage{axodraw}
\usepackage{epsfig}   
\usepackage{amssymb}
\usepackage{latexsym}
\usepackage{amsmath}
\usepackage{graphics,subfigure} 
\usepackage{color}   

\newcommand{\lsp} {\tilde{\chi}_1^0}

\def\chia{\tilde{\chi}_1^0}

\newcommand{\msql} {m_{{\tilde q}_L}}

\newcommand{\mgl} {m_{\tilde g}}

\newcommand{\tanb} {\tan\beta}

\newcommand{\ve} {\nu_e}

\newcommand{\hpm} {h^{\pm}}
\newcommand{\mhpm} {m_{h^{\pm}}}

\def\met        {E\!\!\!\!/_T}

\newcommand{\bea}{\begin{eqnarray}}
\newcommand{\eea}{\end{eqnarray}}
\newcommand{\beq} {\begin{equation}}
\newcommand{\eeq} {\end{equation}}

\begin{document}
%
\title{Prospects for discovering a light charged Higgs boson \\ within the NMSSM  at the FCC-eh collider}

\author{S. P. Das}
\email{ sp.das@uniandes.edu.co}
\affiliation{Department of Physics, Faculty of Science, 
Universidad de los Andes, Apartado A\'ereo 4976-12340, Carrera 1 18A-10, Bloque Ip 
Bogot\'a, Colombia.}
\author{J. Hern\'andez-S\'anchez}
\email{ jaime.hernandez@correo.buap.mx}
\affiliation{Facultad de Ciencias de la Electr\'onica,
  Benem\'erita Universidad Aut\'onoma de Puebla,
\\ Apdo. Postal 542, C.P. 72570 Puebla, Puebla, M\'exico
  and Dual C-P Institute of High Energy Physics, M\'exico.}
\author{S. Moretti}
\email{s.moretti@soton.ac.uk}
\affiliation{School of Physics and Astronomy, University of Southampton, Highfield, Southampton SO17 1BJ, United Kingdom.}
\author{A. Rosado}
\email{rosado@ifuap.buap.mx}
\affiliation{Instituto de F\'{\i}sica, Benem\'erita Universidad Aut\'onoma de Puebla, \\ Apdo. Postal J-48, C.P. 72570 Puebla, Puebla, M\'exico.
\hspace*{2.0truecm}
}
\date{\today}

\begin{abstract}
  We analyze the prospects of observing relatively light charged Higgs bosons ($h^{\pm}$) in their 
  decays via $h^{-} \to s \bar c + s \bar u$ at the upcoming Future Circular Collider in hadron-electron 
  mode (FCC-eh) with $\sqrt s \approx 3.5$ TeV. Assuming that the intermediate Higgs boson ($h_2$) 
  is Standard Model (SM)-like, we study the production of $e^- b \to \nu_e h^{-} b$ (also $b$ could be $\bar b$ in 
  both initial and final states) in the framework of the Next-to-Minimal Supersymmetric Standard Model (NMSSM). 
  We consider constraints from Dark Matter (DM), super-particle and the Higgs boson data. The charged Higgs boson decays 
  into light flavors leads to a three-jets with missing transverse energy signal with one $b$-tagged jet. 
  Our results show that light charged Higgs bosons with mass close to, e.g., 114(121) GeV have the maximal significance 
  of 3.2(1.8)$\sigma$, upon using normal cut based selections and after 1 ab$^{-1}$ of luminosity. However, we further 
  adopt an optimization technique to enhance the latter to 4.4 (2.2)$\sigma$, respectively.
\end{abstract}

\maketitle

\section{Introduction}
\label{sec:intro}
Since the discovery of a Higgs boson, with a mass of 125 GeV,  at the Large Hadron Collider (LHC), by the ATLAS ~\cite{Aad:2012tfa} and CMS 
~\cite{Chatrchyan:2012xdj} experiments,  the Standard Model (SM)  is apparently well established. The current experimental data from ATLAS and CMS \cite{Khachatryan:2016vau} have in fact shown that   this particle is consistent with the prediction of  spontaneous Electro-Weak Symmetry Breaking (EWSB) as  implemented in the SM in its minimal version, which embeds one doublet of Higgs isospin. However, many models with enlarged Higgs sectors still survive these data, because they can well accommodate a SM-like limit. In fact, deviations from SM predictions would be a hint in favor of  new physics in Nature \cite{Ellis:2013lra,*deBlas:2016ojx}. Whereas several new physics scenarios exist that can not only comply with the aforementioned LHC results (as well as explain other experimental observations that cannot be accounted for in the SM, such as neutrino data and Dark Matter (DM))
 but also provide motivated theoretical frameworks (e.g., solving the hierarchy problem of the SM), it is fair to say that Supersymmetry  (SUSY)  is one of the most appealing ones. 

However,
it is very well known that SUSY in its minimal incarnation, called the Minimal Supersymmetric Standard Model  (MSSM) \cite{Kane:1993td,*Gunion:1989we}, has several flaws.
On the theoretical side, 
it suffers from the $\mu$-problem, as this parameter (effectively mixing the SUSY counterparts of Higgs states) ought to be below the TeV scale in order to enable successful EWSB, yet in the MSSM it can really naturally be only zero or close to the Planck mass \cite{Kim:1983dt}. On the experimental side, its allowed parameter space is being more and more constrained from nil searches for new Higgs bosons or Supersymmetric states. Both problems are remedied in the so called Next-to-MSSM (NMSSM) \cite{Ellwanger:2004xm,Barbieri:2013hxa,Miller:2005qua,Ellwanger:2009dp,Maniatis:2009re,Franke:1995tc,Ellwanger:1993hn,*Ellis:1988er,*Derendinger:1983bz,*Nilles:1982dy,Drees:1988fc}, wherein  the Vacuum Expectation Value (VEV) of an additional 
Higgs singlet state  can generate the $\mu $-term at the required  scale and its SUSY counterpart can alleviate experimental bounds as it can act as a new DM state simultaneously altering  SUSY cascade signals and the cosmological relic density. Just like in the MSSM, also the NMSSM has one charged Higgs boson ($h^\pm$) in its spectrum. In fact, a myriad of other non-minimal    SUSY scenarios also have \cite{Khalil:2132388}.

Hence, it is not surprising that  charged Higgs bosons have been the focus of  many searches at the LHC (see, e.g., \cite{Aad:2014kga,*Khachatryan:2015qxa,*Aad:2015typ} for
established analyses and \cite{BergeaasKuutmann:2017yud,*Sirunyan:2017sbn,Laurila:2017phk}
for very recent experimental results), where one normally exploits $h^+\to \tau^+\nu, c\bar s, t\bar b$ (and charged conjugated (c.c.)) decays, which can be searched for model-independently and then interpreted in specific scenarios, like the MSSM or NMSSM  \cite{Akeroyd:2016ymd} \footnote{In fact, also  $h^\pm \to W^\pm Z$ decays have recently been searched for at the LHC \cite{Sirunyan:2017sbn}.}. More recently, the case for studying   the  (non-diagonal) decay $ h^+ \to c \bar{b }$ has also vigorously been made in a variety of new physics scenarios, see \cite{DiazCruz:2009ek,HernandezSanchez:2012eg,Akeroyd:2012yg}, thus  encouraging the 
LHC experimental groups to look for this signal (see, e.g., Ref. \cite{CMS:2016qoa}). It is the purpose of this paper to further 
investigate this last kind of channels, i.e., the non-diagonal ones in flavor space. However, we will do so in other environments than the LHC. Heavier Higgs boson 
within NMSSM has  been recently studied in \cite{Das:2018fog}.

At CERN the future   Large Hadron electron Collider (LHeC) and electron-proton Future Circular Collider (FCC-eh), with center-of-mass energies of 1.3 TeV and 3.5 TeV, respectively \cite{AbelleiraFernandez:2012cc,*Kuze:2018dqd,*Britzger:2017fuc}, offer good prospects as Higgs boson factories, wherein one could elucidate the nature of the couplings of Higgs bosons  to fermions, especially the $h_{\rm SM} \to b \bar{b} $ one, which is difficult to establish at the LHC, but also, e.g., of charged Higgs bosons to generic fermions 
\cite{Das:2015kea,Hernandez-Sanchez:2016vys,Mosomane:2017jcg}.  Given these encouraging results, we specifically analyze here   
the prospects of observing relatively light charged Higgs bosons of the NMSSM decaying 
via   $h^{-} \to s \bar c + s \bar u$. Our work is organized as follows. In section \ref{sec:NMSSM} we describe briefly the NMSSM. Then in  section \ref{sec:NMSSM}  we select some benchmark scenarios for it. In  section \ref{sec:analysis} we give our numerical results whereas in section \ref{sec:conclusion} we finally summarize.

\section{The NMSSM}
\label{sec:NMSSM}
 It is very well known that the NMSSM includes the MSSM Super-fields plus an additional gauge singlet chiral Super-field $\hat{S}$. We focus on the study of the NMSSM as described in the review of Ref.~\cite{Ellwanger:2004xm}, where  $R$-parity and $CP$-conservation are assumed. In such a scenario, as  described in \cite{Das:2017mqw},  the form of the Higgs Super-potential  is
\begin{eqnarray}
W_{\rm Higgs} = (\mu+\lambda \hat{S}) (\hat{H }_u \cdot \hat{H}_d )+\xi_F \hat{S}+ \frac{1}{2} \mu' \hat{S}^2+ \frac{\kappa}{3} \hat{S}^3,
\end{eqnarray}
where $\kappa$, $ \lambda$ are dimensionless Yukawa couplings, the dimensional $\mu$, $\mu'$ parameters are the Supersymmetric mass terms, $\xi_F$  (with dimension two) is the SUSY tadpole term and   the Vacuum Expectation Value (VEV)  $s$ of the singlet $\hat{S}$ generates an effective $\mu$-term (under the assumption that $\mu =0$): 
\begin{eqnarray}
\mu_{\rm eff}=s\lambda.
\end{eqnarray}
Moreover,  the soft SUSY-breaking terms are:
\begin{eqnarray}
- {\cal{L}}_{\rm soft}& =& m_{H_u}^2 |H_u|^2+m_{H_d}^2 |H_d|^2+m_{S}^2 |S|^2+m_{Q}^2 |Q_L^2|+m_{U}^2 |U_R^2|+m_{D}^2 |D_R^2|+m_{L}^2 |L^2|+m_{E}^2 |E_R^2| \\ \nonumber 
&+& (Y_u A_u Q \cdot H_u U_R^c -Y_d A_d Q \cdot H_d D_R^c -Y_e A_e L \cdot H_d E_R^c+ \lambda A_\lambda H_u\cdot H_d S+\frac{1}{3} \kappa A_\kappa S^3 \\
\nonumber 
&+& m_3^2 H_u \cdot H_d + \frac{1}{2} {m'}_s^2 S^2+ \xi_S S + {\rm h.c.}.),
\end{eqnarray}
with all usual parameters of the model ~\cite{Ellwanger:2004xm}. Finally, one can get the Higgs potential  from the Supersymmetric gauge interactions,  soft SUSY-breaking contributions and $F$-term:   
\begin{eqnarray}
V_{\rm Higgs}&=& | \lambda (H_u^+ H^-_d- H_u^0 H^0_d) + \kappa S^2+ \mu' S+|\xi_F |^2 + (m_{H_u}^2+|\lambda S|^2) (|H_u^0|^2 +|H_u^+|^2 ) \nonumber \\
&+&(m_{H_d}^2+|\lambda S|^2) (|H_d^0|^2 +|H_d^+|^2 )+ \frac{g_1^2+g_2^2}{8} \bigg( |H_u^0|^2 +|H_u^+|^2- |H_d^0|^2 -|H_d^-|^2 \bigg)^2 \nonumber \\
&+& \frac{g^2}{2} |H_u^+ H_d^{0*} +H_u^0 H_d^{-*}|^2+m_S^2 |S|^2+ \bigg( \lambda A_\lambda (H^+_u H^-_d-H_u^0 H_d^0)S +\frac{1}{3} \kappa A_\kappa S^3 \nonumber \\
&+& m_3^2 (H_u^+ H_d^- -H_u^0 H_d^0)+\frac{1}{2} m_S'^2 S^2 + \xi_S S+ {\rm h.c.}. \bigg),
\end{eqnarray}
where $g_1$ and $g_2$ are $U(1)_Y$ and $SU(2)_L$ gauge couplings, respectively. In the spectrum of the scalar sector there are seven states: three CP-even neutral Higgs bosons  $h_{1,2,3}$ with $m_{h_1} < m_{h_2}<  m_{h_3}$, two CP-odd scalars $a_{1,2}$ with  $m_{a_1} < m_{a_2}$ and a pair of charged Higgs bosons. As mentioned in \cite{Das:2017mqw}, after spontaneous EWSB, under the convention $\mu=0$, we obtain the model free parameters: $\lambda$, $\kappa$,
$A_\lambda$, $A_\kappa$, $\tan \beta$, $\mu_{\rm eff}$, $m_3^2$, $\mu'$, $m_{S}^{'2}$, $\xi_F$ and $\xi_S$.

We used the package \texttt{NMSSMTools~5.0.1}~\cite{Ellwanger:2004xm} to obtain the SUSY spectrum, i.e., masses and couplings of all NMSSM states. 
We randomly scanned approximately $10^7$ points over the full NMSSM parameter space. The varied parameters and their ranges are 
 given in Tab.~\ref{tab:param} (all the masses and mass parameters in our analysis are in GeV). 
Then, we considered the following  theoretical and experimental constraints \cite{Das:2017mqw}.
\begin{itemize}
\item {Perturbative bounds:} in order to ensure that the NMSSM is perturbative up to a Grand Unified Theory (GUT) scale, it  is necessary to impose that all the points satisfy $\lambda^2+\kappa^2\lesssim 0.7$ \cite{Zheng:2014loa}.    
\item {DM relic density:} in accordance with the Planck measurement \cite{Ade:2015xua}, 
one demands that the relic density for the lightest neutralino must satisfy 
$0.107 < \Omega_{\tilde{\chi}_1^0}<0.131$,  where the NMSSM neutralino is considered as the Weakly Interacting Massive Particle (WIMP)  candidate  for DM assuming the standard cosmological scenario \cite{Drees:2012ji,Sanabria:2014yva}.
\item {Higgs data:} we assume that the intermediate neutral CP-even scalar ($h_2$) is the SM-like Higgs boson, which mass and couplings are constrained by combined studies of ATLAS and CMS \cite{Khachatryan:2016vau} as well as the invisible Branching Ratio (BR) of the SM-like Higgs boson \cite{Khachatryan:2016whc,Aad:2015pla} plus, last but not least, void searches for additional Higgs bosons both at the LHC and previous colliders like LEP/SLD and Tevatron.      
\end{itemize}

\begin{figure}[ht!]
\begin{center}
\raisebox{0.0cm}{\hbox{\includegraphics[angle=0,scale=0.4]{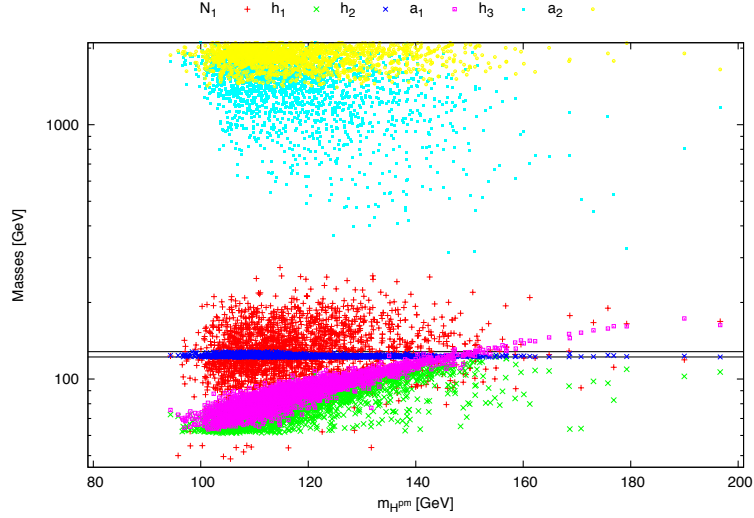}}}
\caption{The masses of the Higgs bosons $h_1$, $h_2$, $h_3$, $a_1$ and $a_2$  and lightest neutralino $N_1$ as a function 
of the charged Higgs mass ($m_{\hpm}$) for the $h_2$-SM scenario (i.e., the $h_2$ is the SM-like state discovered at CERN). 
The masses  of the $h_2$-SM, consistent with coupling ratios and signal strengths following Tab. \ref{tab:coupsigma}, are 
contained within the horizontal lines defining the range [122.1-128.1] GeV and are consistent with all other experimental 
constraints, see details in the text, and so are the other masses too. }
\label{fig:masses}
\end{center}
\end{figure}

\begin{figure}[ht!]
\begin{center}
\raisebox{0.0cm}{\hbox{\includegraphics[angle=0,scale=0.3]{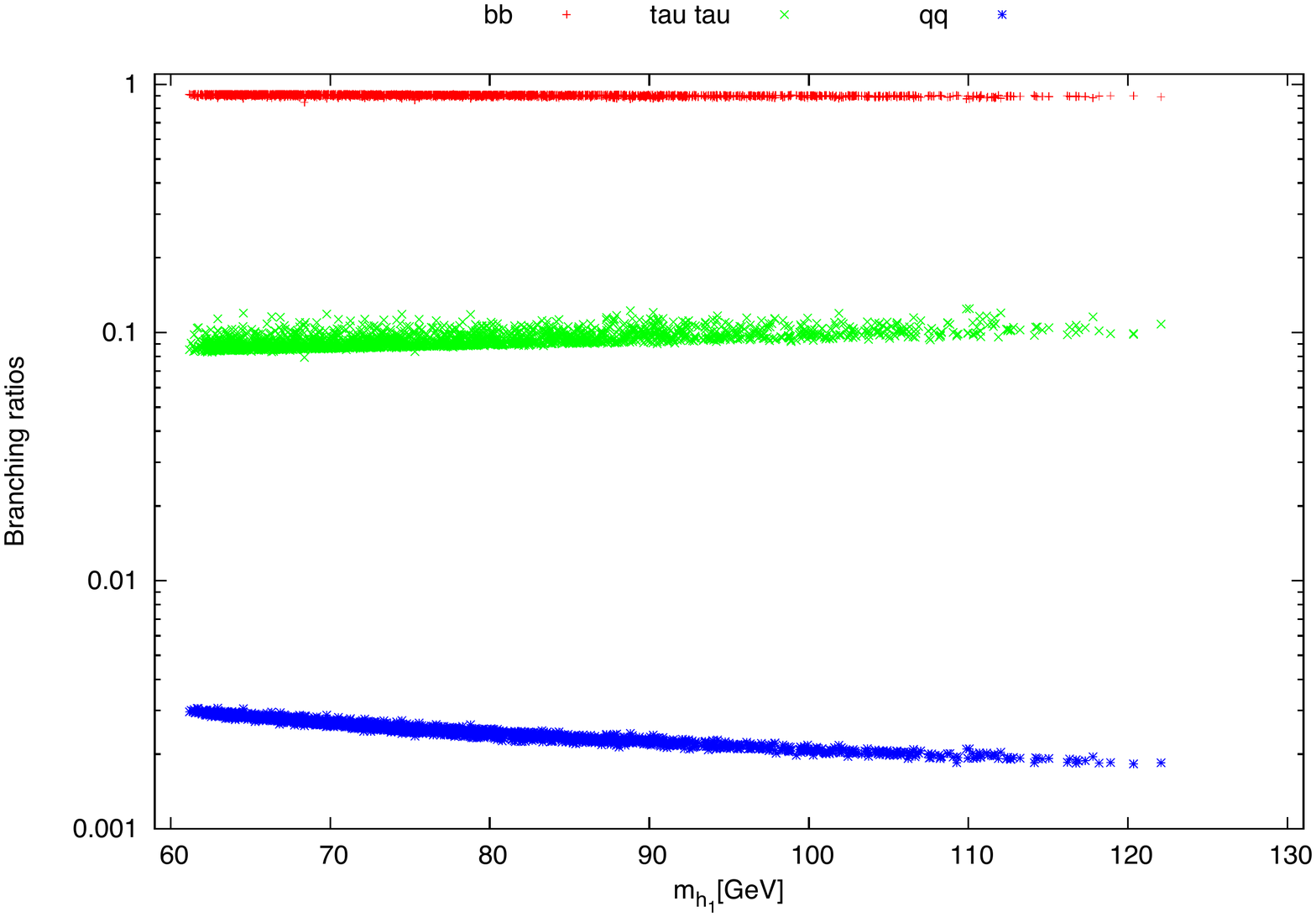}}}
\raisebox{0.0cm}{\hbox{\includegraphics[angle=0,scale=0.3]{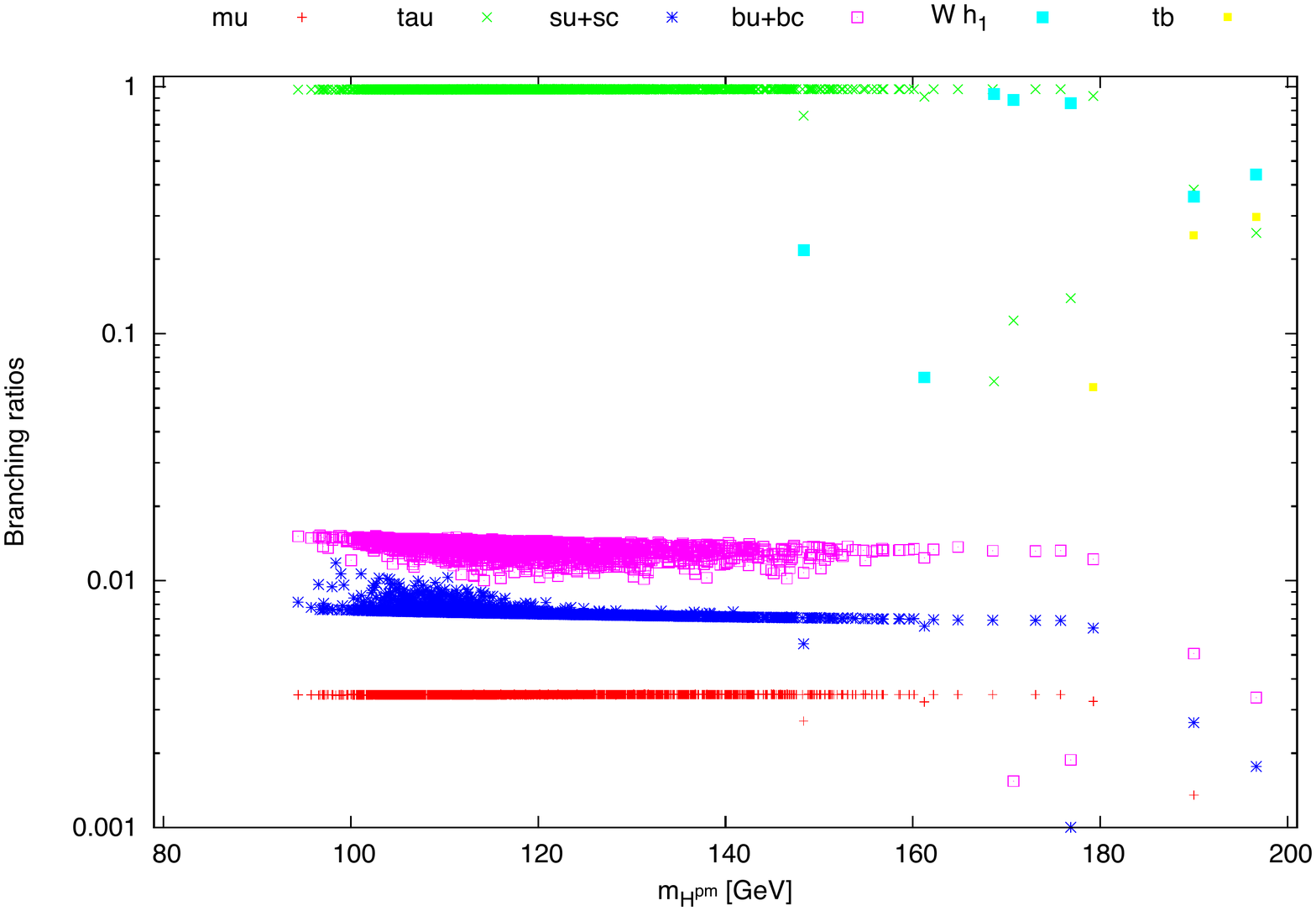}}}
\caption{Left panel: The  three leading BRs of the  lightest (non SM-like) CP-even neutral Higgs state  
$h_1$. Right panel: The BRs of the $\hpm$ state in different channels (with decay rate more than 0.001).}
\label{fig:BRs}
\end{center}
\end{figure}

We are finally left with around 2000 allowed solutions for our  phenomenological 
analysis. Before proceeding to the Signal ($S$) to Background ($B$) analysis aimed at extracting a $h^\pm$ signature, let us study the Higgs boson mass spectrum
(i.e., the masses of $h^\pm,h_1,h_2,h_3,a_1$ and $a_2$  and as well the Branching Ratios (BRs) of the $h^\pm$ and $h_1$ states (as the former can decay into the latter over the region of NMSSM parameter space that we have scanned). This is done in Figs.~\ref{fig:masses} and ~\ref{fig:BRs}, respectively.  From these plots, it is clear that there exist regions of the NMSSM parameter space wherein the $\hpm$ state can be sufficiently light to be copiously produced through the aforementioned $e^-p$ channel while decaying sizably through the  fermionic channels that we are seeking to establish. 

\begin{table}[t!]
\centering
{\scriptsize
\begin{tabular}{||c||c|c||}
\hline
Parameters&Min&Max\\
\hline
\hline
$\lambda$&0.001& 0.7\\
$\kappa$& 0.001& 0.7\\
$A_{\lambda}$&100.0& 2500.0\\
$A_{\kappa}$& -2500.0& 100.0\\
$\tanb$&1.5 & 60.0 \\
$\mu_{\rm eff}$& 100.0& 500.0\\
$M_1$& 50.0& 400.0\\
$M_2$& 50.0& 500.0\\
$\msql$& 300.0& 1500.0\\
$A_t$=$A_b$& -4000.0& 1000.0\\
$M_A$& 100.0& 500.0\\
$M_P$& 100.0& 3000.0\\
\hline
\hline
\end{tabular}
}
\caption{The minimum and maximum values of the varied NMSSM parameters. The following inputs  
remain fixed: $M_3$ = 1900.0 GeV (this allows the gluino mass $\mgl$ to be above the mass 
limits from recent LHC Run 2 limits); $m_{\tilde \ell}$ = 350.0 GeV (for all three generation as 
well as left- and right-handed states) and $A_{\tau}$=$A_e$=$A_{\mu}$ = 1500.0 GeV. Here, $M_A$($M_P$) 
is the doublet(singlet) component of the CP-odd Higgs mass matrix. Besides,  masses of the right handed squark are assumed
same as masses of the left-handed quarks  
}
\label{tab:param}
\end{table}

\begin{table}[t!]
\centering
{\scriptsize
\begin{tabular}{||c||c|c||c||c|c||}
\hline
Parameters&Min&Max & Parameters&Min&Max \\
\hline
\hline
$\kappa_W$&0.81&0.99 & $\mu_{VBF}^{\tau \tau}$&0.50&2.10 \\
$\kappa_{t}$&0.99&1.89 & $\mu_{ggF}^{\tau \tau}$&-0.20&2.20\\
$|\kappa_{\gamma}|$&0.72&1.10 & $\mu_{VH}^{bb}$&0.00&2.00\\
$|\kappa_{g}|$&0.61&1.07 & $\mu_{ttH}^{bb}$&-0.90&3.10\\
$|\kappa_{\tau}|$&0.65&1.11 & $\mu_{VBF}^{WW}$&0.40&2.00\\
$|\kappa_{b}|$&0.25&0.89 &$\mu_{ggF}^{ZZ}$&0.51&1.81\\
BR$(h_2 \to {\rm inv})$& & 0.25 &$\mu_{VBF}^{\gamma\gamma}$&0.30&2.30\\
&& &$\mu_{ggF}^{\gamma\gamma}$&0.66&1.56\\
\hline
\hline
\end{tabular}
}
\caption{The normalised couplings ($\kappa$) and signal strength ($\mu$) of the SM-like Higgs state $h_2$ have been allowed to vary within 2$\sigma$ ranges from the combined ATLAS and CMS measurements of Ref. \cite{Khachatryan:2016vau}, following Tabs. 17 (upper panel) 
and 8 herein, respectively.}
\label{tab:coupsigma}
\end{table}

\begin{table}[!ht]
\vspace*{-5mm}
\vspace{3mm}
\footnotesize
\begin{center}
\scalebox{0.9}{
\begin{tabular}{|l||r|r|r||}
\hline
{Benchmark Points (BP)}&1&2&3
\\\hline
$\lambda$& 0.3230& 0.1859& 0.0803
\\\hline
$\kappa$& 0.0103&  0.0218&0.0595
\\\hline
$\tan \beta$& 2.95 & 58.30& 53.06
\\\hline
$A_\lambda$(GeV)& 1521.5&  500.8& 852.8
\\\hline
$A_\kappa$(GeV)& -1056.2&  924.7& -772.4
\\\hline
$\mu_{\rm eff}$(GeV)& 372.14& 386.99&  475.31
\\\hline
$M_1$ (GeV)& 129.13& 196.64& 173.49
\\\hline 
$M_2$ (GeV)& 231.37& 258.92& 311.00
\\\hline
$M_{\tilde q}$ (GeV)& 671.4& 687.5& 751.1
\\\hline
$A_t$ = $A_b$ (GeV)& -1687.1& -1680.8& -1888.9
\\\hline
$M_A$ (GeV)& 94.2& 243.3& 77.1
\\\hline
$M_P$ (GeV)& 1885.3& 1839.8& 1823.4
\\\hline\hline
$\xi_F$ ($10^6$$GeV^2$)& -1.76& -1.13& -7.12
\\\hline\hline
$\xi_S$ ($10^9$$GeV^3$)& -3.97& -7.10& -4.83
\\\hline\hline
$m_{h_1}$ (GeV)&  68.39&  63.95& 71.23
\\\hline
$m_{h_2}$ (GeV)& 127.96& 122.76& 122.79
\\\hline
$m_{h_3}$ (GeV)& 1851.57& 1860.38& 1020.22
\\\hline
$m_{a_1}$ (GeV)&  69.73& 67.35&  77.07
\\\hline
$m_{a_2}$ (GeV)& 1884.33& 1839.33& 1823.36
\\\hline
$m_{h^\pm}$ (GeV)&98.41 & 114.63&121.27
\\\hline\hline
$m_{\chia}$ (GeV)& 27.65& 89.58& 169.53
\\\hline
BR($\hpm \to b \bar c + b \bar u$ + c.c.)& 0.0149& 0.0129&0.0115
\\\hline
BR($\hpm \to s \bar c + s \bar u$ + c.c.)& 0.0118& 0.0074& 0.0073
\\\hline
$\sigma$ [fb]& 932.57&4325.76& 2592.28
\\\hline
Factor& 0.0118&0.0074& 0.0073
\\\hline
$\sigma$.Factor[fb]& 1098.84 &  3200.52& 1900.58
\\\hline\hline
\end{tabular}
}
\end{center}
\caption{The selected NMSSM benchmark points obtained using \texttt{NMSSMTools~5.0.1}~\cite{Ellwanger:2004xm} 
to find the maximal event rates for four different signal  at the FCC-eh for $\sqrt s \approx 3.5$ TeV. The values displayed are 
at the EW scale. The following parameters are fixed: $M_3$ =1900 GeV, $A_{\tau}$ = $A_{\ell}$=1500 GeV 
and $M_{\tilde \ell}$ = 350 GeV.  We use $M_A$ and $M_P$ as inputs, thus our scenario is 
not the $Z_3$-NMSSM. Further,  $\xi_F$ and $\xi_S$ are non-zero and also given in the table. 
The factor represents the BR of the charged Higgs decay under  study, i.e.,  $h^{-} \to s \bar{c} + s \bar{u}$. 
We select the best three BPs for which the event rates are maximal in this decay channel.}
\label{table:nmssmbp}
\end{table}

\section{Numerical Analysis}
\label{sec:analysis}
\begin{figure}[ht!]
\begin{center}
\includegraphics[angle=0,scale=0.25]{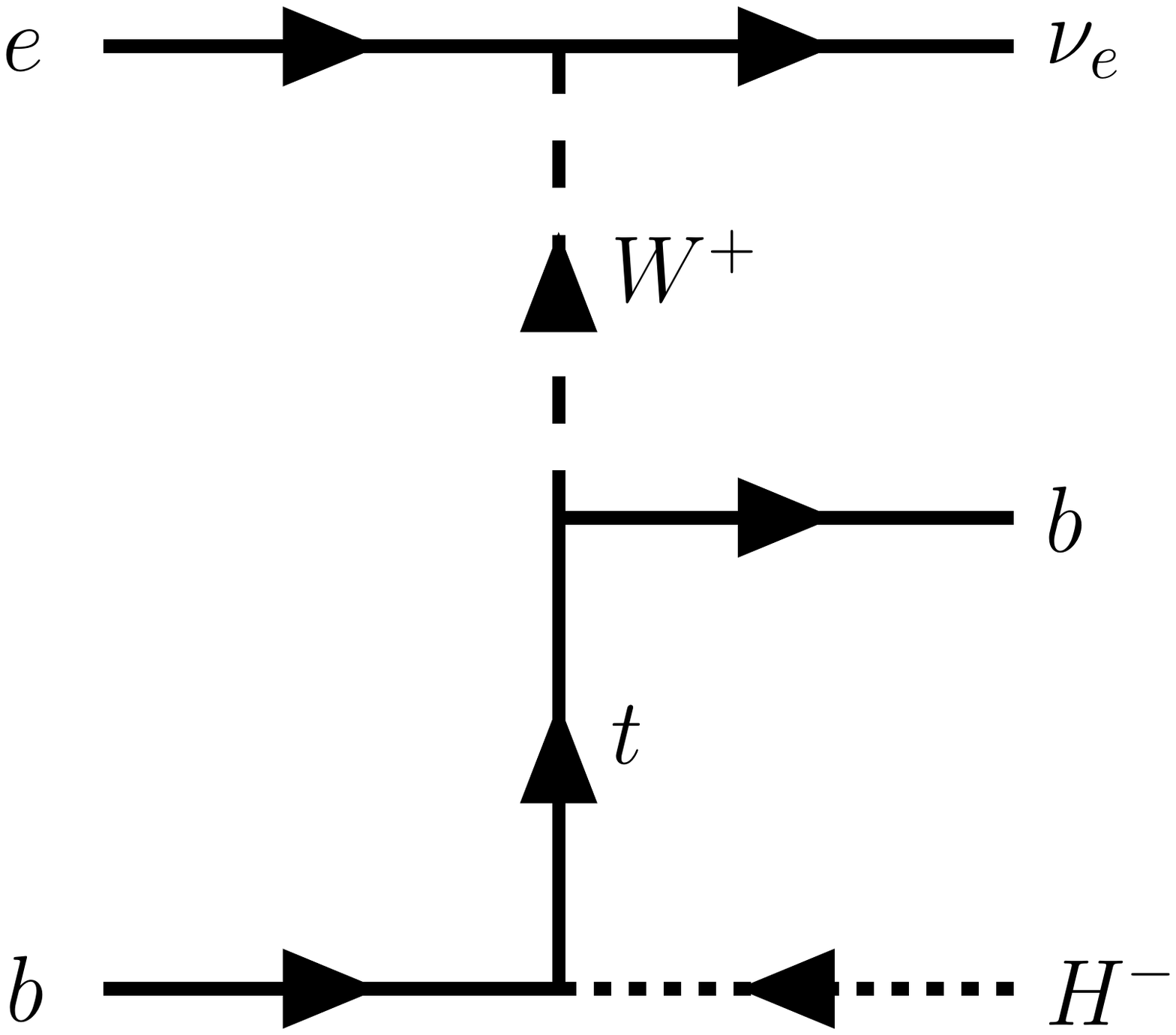}
\caption{Representative Feynman diagram for the subprocess $e^- b \to \nu_e h^{-} b$. In the $t$-channel all other possibilities 
with neutral CP-even and CP-odd Higgs bosons within the NMSSM have been considered in the hard process under consideration. 
Also the sea-quark fluxes from the proton beam (e.g., for the process $e^- \bar b \to \nu_e h^{-} \bar b$) have  been  considered.}
\label{fig:diagrams}
\end{center}
\end{figure}

The production process that we are considering is  $e^- p \to \nu \hpm b$ (and also the $\bar{ b}$ in both initial and final state),  the proton having a $b$ parton flux necessary 
to enable the process. The signal diagrams are sketched in Fig.~\ref{fig:diagrams}. All of these are expected to produce 
a $b$-jet in the final state with enough transverse momentum to enter the detector region. As mentioned, the $h^{-}$ decays 
searched for are into $s \bar c + s\bar u$ final states, so that the largest irreducible background is given by similar 
Feynman graphs where $h^\pm$ is replaced by $W^\pm$. Several irreducible backgrounds will also be considered.

We start by showing some inclusive $h^\pm$ production and decay rates, by categorizing them depending on the detector signatures that they originate.   They are the following ones (here, $j$ refers to both a light $q=d,u,s,c,g$ or heavy $b$ quark while $\ell=e,\mu$), wherein (for the time being) we overlook describing possible jet-tagging procedures of the hadronic components of the signals.

{Decay cascade A}: $h^{-} \to b\bar u + b \bar c$ yielding {3j + $\met$}.
 
{Decay cascade B}:  $h^{-} \to s\bar u + s \bar c$  yielding {3j + $\met$}.

{Decay cascade C}: $h^{-} \to W^{-} h_1 \to q \bar q' b \bar b$ yielding {5j + $\met$}.

{Decay cascade D}: $h^{-} \to W^{-} h_1 \to \bar\ell {\nu_{\ell}} b \bar b$ yielding {3j +1$\ell$ + $\met$}. 

\begin{figure}[ht!]
\begin{center}
\raisebox{0.0cm}{\hbox{\includegraphics[angle=0,scale=0.5]{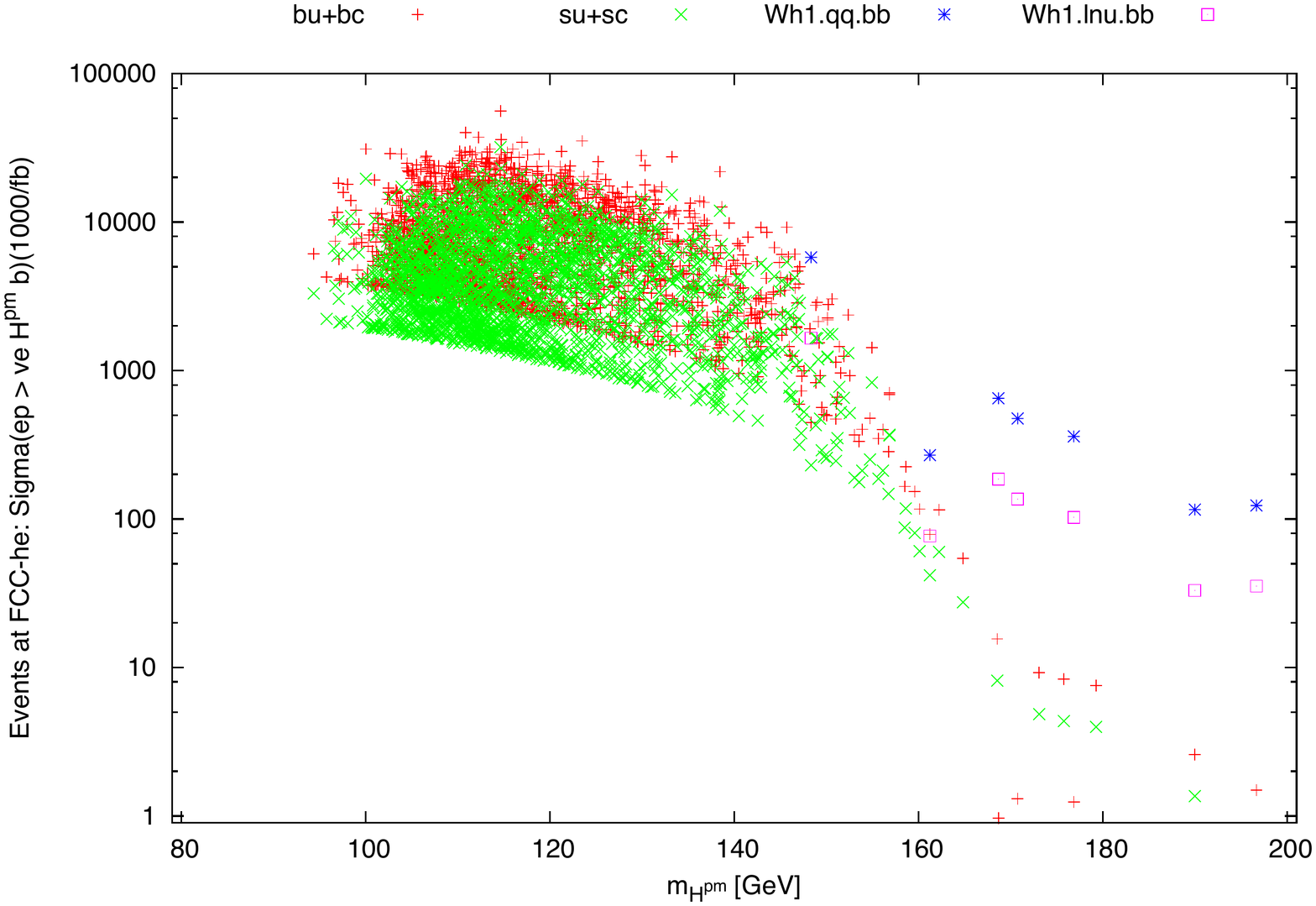}}}
\caption{Number of events from $e^{-} b \to \nu_e h^{-} b$ (plus $e^- \bar  b  \to \nu_e h^{-} \bar b$) in different decay cascades of the charged Higgs bosons after an integrated luminosity of 1 ab$^{-1}$ 
with the inclusion of $W^{\pm}$ BRs  in both hadronic ($jj$) and leptonic ($\ell \nu_{\ell}$, $\ell$ = $e$, $\mu$) modes.}
\label{fig:eventrates}
\end{center}
\end{figure}

From the distributions of the corresponding event rates (taken, e.g., at 1 ab$^{-1}$) seen in Fig.~\ref{fig:eventrates}, it is clear that the fermionic decays (A and B) can dominate over those involving the $W^-$ and lightest neutral Higgs boson $h_1$ (C and D). As case 
A has already been dealt with in Ref.~\cite{Akeroyd:2012yg}, in this article, we study the feasibility of finding the charged Higgs 
signal in the 3j + $\met$ channel of case B, while  we are neglecting considering further  options C and D. (The last two cases 
will be addressed in a forthcoming report \cite{rhmd}.) In order to pursue its study, we have selected three NMSSM representative 
Benchmark Points (BPs), each maximizing(minimizing) $h^-$ fermionic(bosonic) decays, for which we have tabulated the 
properties in Tab. \ref{table:nmssmbp}, wherein cross section rates are subjects to the cuts of Eq.(\ref{presel}). We will 
perform a $S$-to-$B$ analysis for these three selected BPs.
\subsection{Higgs boson signals}
\label{sec:higgsignal}
We study the  scenario where the $h_2$ state is the SM-like neutral Higgs boson discovered at CERN  in 2012. The production process of the charged Higgs boson is $e^-p\to \nu_e h^- b$ and the Higgs boson production 
in our analysis is mainly dominated by the pair of diagrams we have sketched. 
We have estimated the parton level signal cross sections using \texttt{MadGraph v2.4.3} \cite{Alwall:2014hca}.  
The allowed NMSSM model parameter space from \texttt{NMSSMTools~5.0.1}~\cite{Ellwanger:2004xm} are
written in SLHA format and fed to \texttt{MadGraph v2.4.3}. The BRs
of the Higgs boson in all the decay modes are estimated by using \texttt{NMHDECAY}~\cite{Ellwanger:2004xm}.
To obtain the cross sections at the FCC-eh \cite{Bruening:2013bga,AbelleiraFernandez:2012ty,
AbelleiraFernandez:2012cc,Appleby:2013sha}, we consider an electron beam of energy $E_{e^-}$= 60 GeV and
a proton beam of energy $E_{p}$= 50000 GeV, corresponding to a Center-of-Mass (CM) energy of
approximately $\sqrt s = 3.5$ TeV. To estimate the signal event rates at parton level we applied the following 
basic pre-selections:  
\beq \label{presel}
p^{q,e}_T >  10~{\rm GeV},  \qquad \eta^{q,e} < 5.5, \qquad  \Delta R (qq,qe) > 0.2,
\eeq
 where ($q$=$u,d,c,s,b,g$ ) and $\Delta R^2 = \Delta \eta^2 + \Delta \phi^2$, here $\eta$ and $\phi$ are 
the pseudo-rapidity and azimuthal angle, respectively. We take $m_t$=173.3 GeV as the top-quark (pole) mass.

We have set the renormalization and factorization scales at $\sqrt {\hat s}$, the CM energy 
at the parton level, and adopted the NNPDF23LO Parton Distribution Functions (PDFs) \cite{Ball:2013hta,Pumplin:2002vw}  
 with $\alpha_{\rm s}$ (the strong coupling constant within the four-flavor scheme) evaluated  
consistently at all stages (i.e., convoluting PDFs, hard scattering and decays). Parton shower (both initial 
and final), hadronization, heavy hadron decays, etc. have been dealt with by {\tt PYTHIA v.6.428} \cite{Sjostrand:2006za}.
We consider all the light-flavor quarks, $b$-quark and gluon in the proton flux. Any flavor-mixing, wherever 
appropriate, is also considered for the allowed diagrams.
We  also note  that a final state forward jet could also be a $b$-jet. However, as it is mostly in the  
forward region, with the tight constraints in rapidity of a $b$-taggable jet, it hardly
qualifies as $b$-tagged jet.

\subsection{Backgrounds}
\label{sec:backg}

There are mainly two groups of SM backgrounds to our Higgs signal. The charged-current backgrounds 
consisting of $\nu t \bar b$, $\nu b \bar b j$, $\nu b2j$, $\nu 3j$ and the neutral-current  
ones identified as $e^{-} b \bar b j$, $e^{-} t \bar t$,  $e^{-} b jj$ and $e^{-} jjj$. In all of 
these backgrounds the charge-conjugated processes are naturally implied, like for the signal. We generated also the SM backgrounds at the parton level using \texttt{MadGraph v2.4.3} \cite{Alwall:2014hca}
and then fed them to {\tt PYTHIA} v.6.428 \cite{Sjostrand:2006za} for parton showering
(both initial and final), hadronization, heavy hadron decays etc. 
 The expected background rates  in [pb], after the aforementioned pre-selections,   
are given in the fourth column of Tab.  \ref{tab:smbgs}.   

\begin{table}[h!]
\begin{center}
\begin{tabular}{|c|c|c|c|}
\hline
Label& Final state & $j$ &FCC-eh[pb]\\
\hline
\hline
p1 & $\nu bbj$ &  $u,d,s,c,g$ & 0.537\\
p2 & $\nu bjj$ & $u,d,s,c,g$ & 0.491\\
p3 & $\nu bt$ & -- &  8.14\\ 
p4 & $\nu jjj$ &  $u,d,s,c,g$ & 137.37\\
\hline\hline
p5 & $ebbj$ & $u,d,s,c,g$ & 106.13\\
p6 & $ebjj$ & $u,d,s,c,g$ & 15.66\\
p7 & $ett$ & --  & 0.395\\
p8 & $ejjj$ &  $u,d,s,c,g$ & 1330.81 \\
\hline
\end{tabular}
\caption{List of background processes via charged (top, p1--p4) and neutral (bottom, p5--p8) currents with their total cross sections in [pb] at the FCC-eh with $E_{e^-}$=60 
GeV with  $E_p=50$ TeV after the pre-selection cuts (\ref{presel}). Charge conjugate final states 
are included. The symbol $j$ stands for different partons, as listed in the third column, 
where the corresponding antiquarks are always included. See the text for further explanations. 
(Note that a lepton veto will eventually reduce p5--p8 to a large extent, thus the main 
irreducible backgrounds for the $3j +\met$ channel will be  p1--p4.)}
\label{tab:smbgs}
\end{center}
\end{table}
 
\subsection{$S$-to-$B$ analysis}
\label{sec:sigbganalysis}

The Initial State Radiation (ISR) is included and will reduce the total CM energy of the collision, however, 
at the FCC-eh, with the main dynamics along the $t$-channel, the effective CM energy loss due to ISR has a 
reduced impact. In contrast, the  four-momenta of the jets are different as compared to the parton level
quark ones due to Final State Radiation (FSR), so that in our analysis we have considered Gaussian smearing 
effects from the FCC-eh detectors and their parameters are treated similarly to what done in our recent 
analysis for  the  LHeC \cite{Das:2015kea,Das:2016eob}. However, to be complete, let us describe these 
detector aspects here, albeit briefly.

The toy calorimeter {\tt PYCELL} adapted to the LHeC detector parameters is considered as reference \cite{Das:2016eob}.  We 
apply a symmetric and large rapidity coverage for both jets and leptons (hereafter, we consider only electrons): in accordance 
with LHeC detectors  \cite{AbelleiraFernandez:2012cc}, we take $|\eta^{j,e}| < 5.5$ as coverage, with segmentation
$\Delta \eta \times \Delta \phi = 0.0359 \times 0.0314$  (the number of division in $\eta$ and $\phi$ are 320 and 200, 
respectively). Besides, the Gaussian energy resolution of \cite{Bruening:2013bga} for electrons ($e$) and jets ($j$) is 
adopted here. A cone algorithm for  jet-finding is also used, with jet radius $\Delta R= \sqrt{\Delta\eta^{2}+\Delta\phi^{2}} = 0.5$. Finally, the selection and isolation criterion for jets and electrons are the same as in \cite{Das:2015kea,Das:2016eob}.  Leptons ($e$, $\mu$) are selected when satisfy the requirements: $E_T^\ell \geq 15$ GeV and $|\eta^\ell | \leq 3.0$. We 
have included leptons as parts of jets and the isolation criterion among these is $\Delta R (j , \ell) \leq 0.5$, with two 
situations: {a) when the $E_T$ of a jet is close to a lepton one, namely, $0.8 \leq E^j_T/E^\ell_T \leq 1.2$, this particular jet is 
removed from the list of original jets and treated as a lepton and b) when the $E_T$ of the jet instead differs substantially from the lepton one, the 
latter is removed from the list of original leptons. This isolation criterion mostly removes leptons from $b$ or $c$ decays.}
Further, jets with $E_T^j  \geq 15$ GeV and $|\eta^j | \leq 2.5$ with $ b(c)$-flavored hadron (B(C)-hadron) with 
$\Delta R (j, B(C)) \leq 0.2$ is considered  taggable, with a probability $\epsilon_{b(c)} = 0.5(0.1)$. For the other 
jets ($j=u,d,s,g$) the probability is $\epsilon_j=0.01$. Also note that the top quark and $W^\pm$ boson were 
allowed to decay freely within the {\tt PYTHIA} program.

\subsection{The $3j$ $+$ $\met$ channel}
\label{sec:sigb}

In this subsection we will analyze our charged current signal yielding a 3$j$ $+$ $\met$ signature and apply  
different kinematical  selection cuts to isolate it from the backgrounds. We describe each of out selections individually.

\begin{itemize}

\item {\bf Cut A:} {$N_{\rm jet} \ge 3$ and varying $p_T^j$ thresholds}. We first selected events containing at least three jets, i.e., $N_{\rm jet} \ge 3$. 
The transverse momentum requirements on the jets are varied for the three signal BPs. We applied 
$p_T^j > $ 15.0 GeV, 20.0 GeV and 25.0 GeV for the following masses  of the charged Higgs boson: 98.4 GeV, 114.6 GeV and 121.3 GeV, respectively. 
All  them with  identical rapidity coverage of $\eta^j < 5.5$. The number of jet 
($N_{\rm jet}$) distribution is shown in the left panel of Fig. \ref{njetbtag}. The signal efficiencies for having
$N_{\rm jet} \ge 3$ are, approximately, 70.1\%, 58.4\% and 44.6\%, respectively \footnote{The 
efficiencies quoted in these numerical sections are always with respect to the previous set of selections.}. Amongst all 
backgrounds, $e t \bar t$ leads to a total of six-jets if both top quarks decay hadronically: here, the jet efficiency 
is maximal, in the range of 94.0 to 88.0\% (for BP1 to BP3, respectively). The next highest efficiency is from 
$\nu t b$, where the maximal number of jets is four. This cut leads here to an efficiency around 67.2 to 49.3 \% (for BP1 to BP3, respectively). The jet efficiency  
for the $\nu bbj$ (26.4 - 14.9\%), $\nu bjj$ (30.8 - 16.1\%) and $\nu jjj$ (27.5 - 14.1\%) (charged current) noises are worse whereas for the neutral current ones $ebbj$ (10.2 - 4.1\%), $ebjj$ (12.6 - 4.1\%) and $ejjj$ (12.8 - 4.6\%) they are even lower. This is due to the fact 
that, in the former case, there is no (and there cannot be as it escapes detection) strict 
selection of neutrino momentum whereas, in the latter case,  a minimum transverse momentum ({on lepton}) of around 5 GeV 
is imposed by detector requirements. Therefore,  even if, e.g.,  the $\nu bbj$ and $ebbj$ processes have rather identical 
properties from the jet perspective, the efficiencies in $ebbj$ is lower than the corresponding one in $\nu bbj$ and this holds for the other two cases too. 
Further, the aforementioned lepton-jet isolation criterion also plays a role in reducing the number of jets in the backgrounds 
with an explicit lepton. However,  a nearly complete elimination of these channels is achieved by the next cut. 
\begin{figure}[ht!]
\begin{center}
\raisebox{0.0cm}{\hbox{\includegraphics[angle=0,scale=0.42]{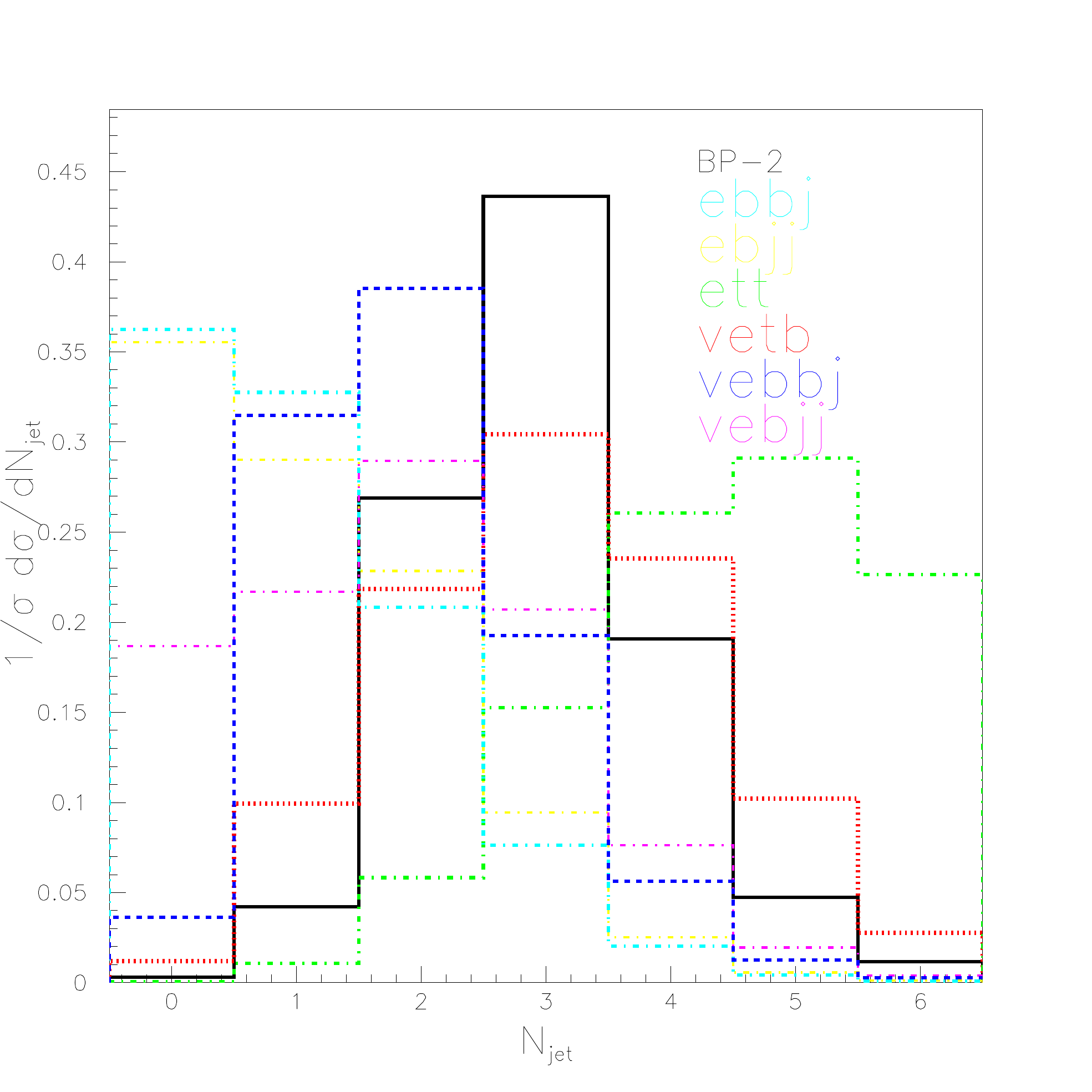}}}
\raisebox{0.0cm}{\hbox{\includegraphics[angle=0,scale=0.42]{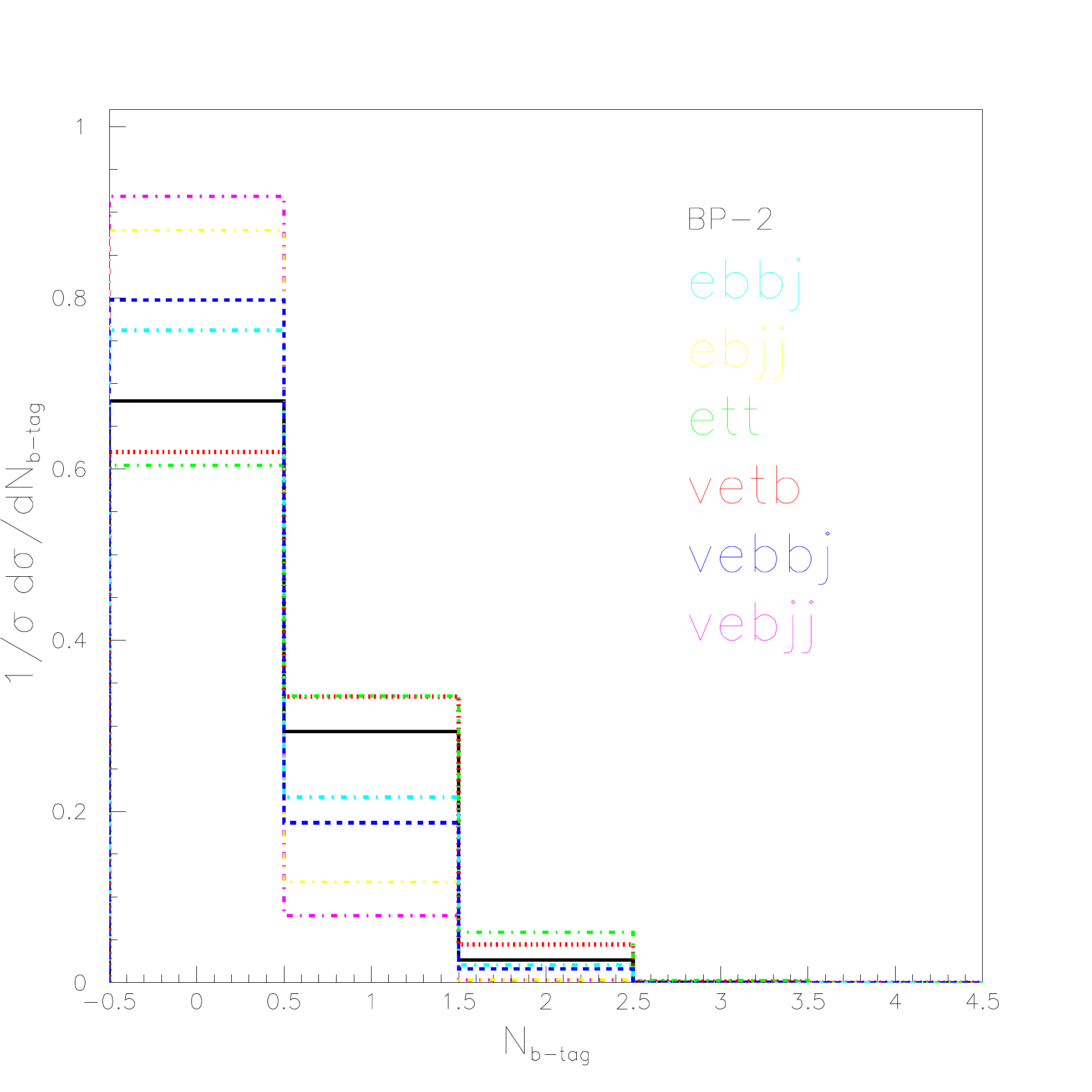}}}
\caption{Left panel: Number of jet ($N_{\rm jet}$) distribution for BP2 (it is very similar for BP1 and BP3) as well as the 
SM backgrounds. 
Right panel: The number of $b$-tagged jet ($N_b$) distribution  for  BP2 (it is very similar for BP1 and BP3) 
as well as the SM backgrounds (herein, the  distributions for $\ve jjj$ and $ejjj$ are having higher entries 
in the left-most bin, i.e., no-tagged jets at all). For both observables, the distribution for $\ve jjj$($ejjj$) is very similar to that for $\ve bjj$($ebjj$). }
\label{njetbtag}
\end{center}
\end{figure}
\item {\bf Cut B:}{ $N_{\rm lep}=0$, the lepton veto.} We required no presence of any lepton (here, electron) in our events. 
The distribution of number of leptons is shown in the left panel of Fig. \ref{nlepmet}. If we find an electron 
with $p_T^e$ $>$ 15.0 GeV and $\eta^e$ $<$ 3.0, we reject such configurations. That is, we apply a lepton veto.
The signals survive at a rate of approximately 93.2\%, 95.6\% and 94.7\%, respectively, for BP1, BP2 and BP3. As intimated,  
this lepton veto criterion largely reduces the SM backgrounds that contain an explicit electron.
For example, $ebbj$, $ebjj$ and $ejjj$ survive at a rate  (from BP1 to  BP3)  in the 
ranges 20.0 - 17.0\%,  21.9 - 17.7\% and 24.4 - 21.3\%, respectively. For $ett$ the 
survival efficiency is close to 8.5\% in all three benchmarks, since all top quark decays (including via $\tau$'s) are enabled. 
Somewhat similar effects happen for $\nu tb$ too, with approximately 15.0\% of events surviving 
in the case of all three benchmarks. Furthermore, there are also secondary sources of electrons, like semi-leptonic 
$b$ decays or prompt meson decays. Taking this into account, the transverse momentum and rapidity criterion reduces the efficiency by approximately a further 5\%.
The efficiencies for $\nu bbj$ is approximately 6.5\%, where the source of the lepton
is  from a  semi-leptonic $b$ decay or from secondary sources
like meson decay or photon mis-identification. The efficiency for $\nu bjj$, being 2.8\%, is just half of that for $\nu bbj$,  
as it is clear from the relative presence of $b$-jets in these two cases. In case of $\nu jjj$, 
the lepton would only be coming from  secondary sources (meson or photon) during fragmentation  
and hadronization, so that the efficiency to  have one isolated lepton is approximately 1.0\% on average  
with mild changes across the three benchmarks.
One can note that we have not considered here the 
lepton mis-tagging efficiency from the jets. This is approximately 0.001\% and having the three (or more) 
jets explicitly after considering the ISR and FSR, this efficiencies are somewhat consistent 
with the mis-tagging numbers with proper combinatorics. 
\begin{figure}[ht!]
\begin{center}
\raisebox{0.0cm}{\hbox{\includegraphics[angle=0,scale=0.42]{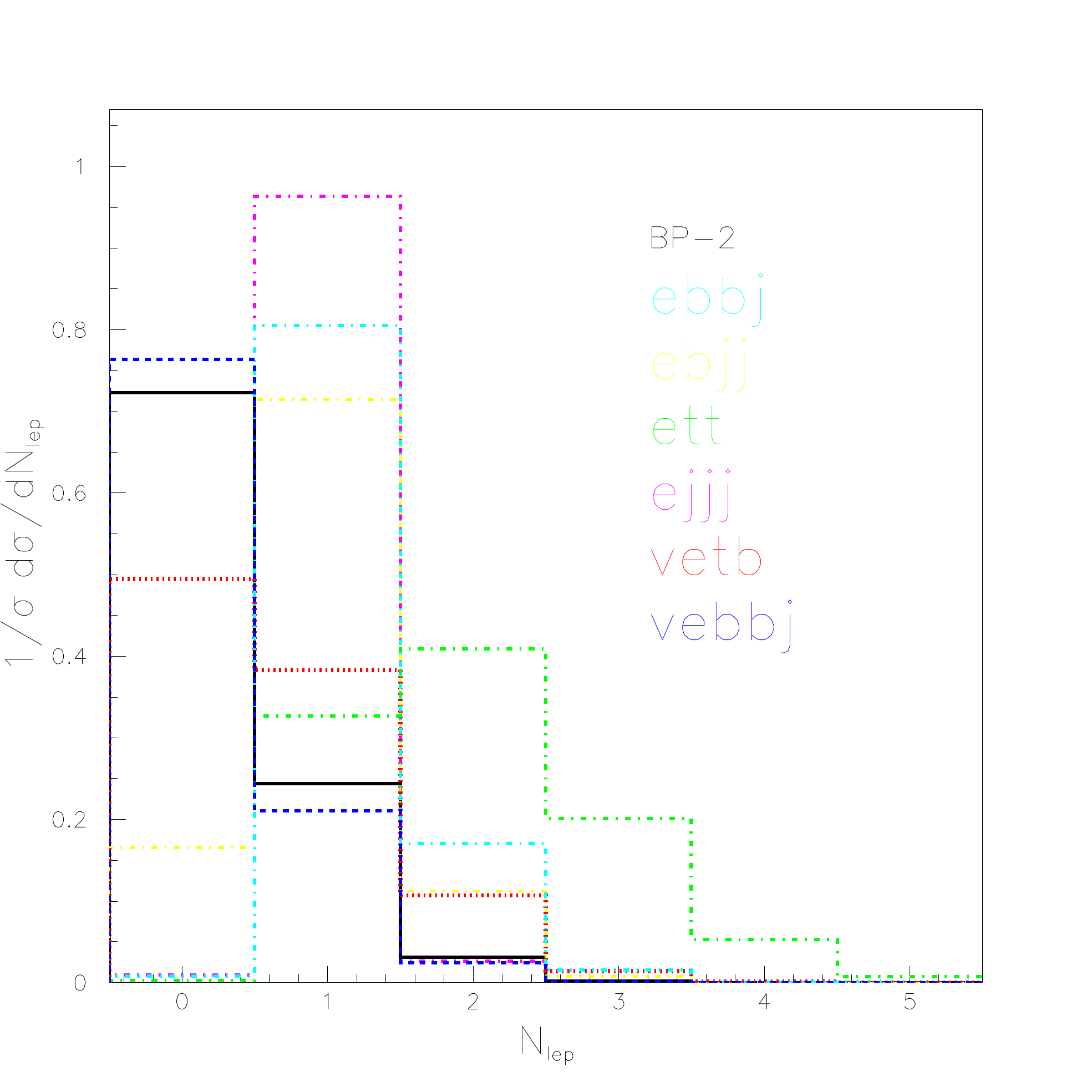}}}
\raisebox{0.0cm}{\hbox{\includegraphics[angle=0,scale=0.42]{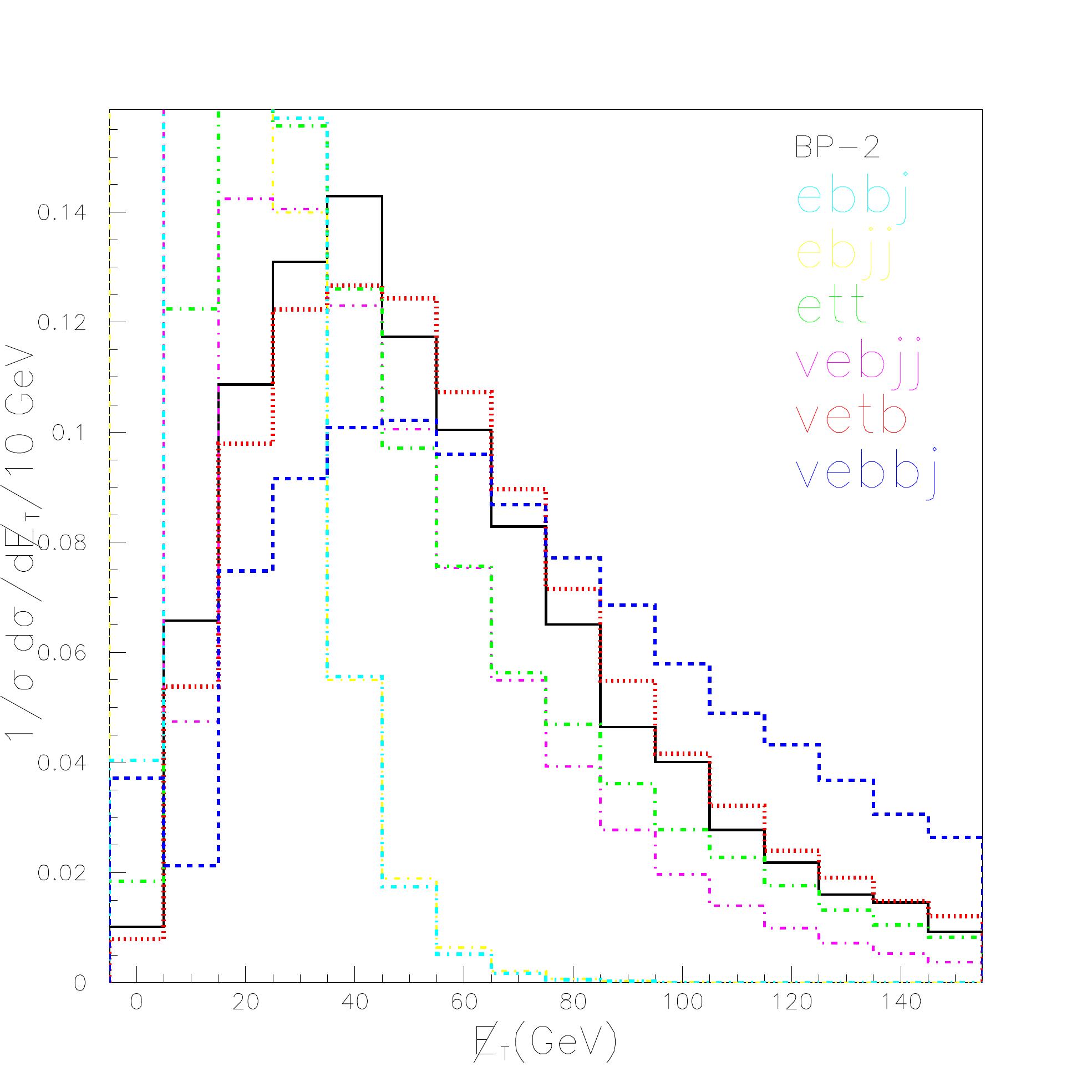}}}
\caption{Left panel: Number of electron ($N_e$) distribution for  BP2 (it is very similar for BP1 and BP3) 
as well as the SM backgrounds. Right panel: Missing transverse energy ($\met$) distribution for BP2 (it is 
very similar for BP1 and BP3) as well as the SM backgrounds. For both observables, the distribution for 
$\nu jjj$($ejjj$) is very similar to that for $\nu bjj$($ebjj$).}
\label{nlepmet}
\end{center}
\end{figure}
\item {\bf Cut C:} {$\met$ $\ge$ 20.0 GeV.} The signal contains a neutrino and this explicitly leads 
to missing transverse energy (other than what coning from jet smearing and mis-measurements). The distribution of 
missing energy is shown in the right panel of Fig.~\ref{nlepmet}. We demanded that $\met$ should be 
larger than 20 GeV. All signal BPs survived at the level of  87.5 - 88.4\%. The background $\nu bbj$ 
survived approximately 98\% while $\nu bjj$($\nu jjj$)  survived around 91 - 93\%(94 - 96\%) 
for the three BPs. The $\nu tb$ noise survived approximately 90\% for all  benchmarks.
The $ebjj$  survival rate is  30\% - 44\% while for $ebbj$ has 37.3\% - 49.8\% for all the benchmarks. The 
relatively larger survival rate for $ebbj$ is due to the semi-leptonic decays of the two bottom quarks 
which produce neutrinos. The $ejjj$ survival rate,  27.7\% - 43.7\%,  is little less than the $ebjj$ one 
as no sources of neutrinos from  $b$-quark semileptonic decays exist but only the jet mis-measurements. For $ett$, 
the efficiencies are 72.6 - 78.5\%, the large survival fractions being due to the leptonic decays 
of the top quarks,  the jet mis-measurements or a combinations thereof in case of top quark mixed decays.

\item {\bf Cut D: $N_{b}$ $= 1$.} We demanded exactly one $b$-tagged jet with 
the inclusion of proper mis-tagging. The distributions of the number of $b$-tagged 
jets ($N_{b}$) are shown in the right panel of Fig. \ref{njetbtag}. It is clear 
from the spectra given  that the signal mostly contains only one $b$-tagged jet. If the two jets stemming from the $h^\pm$ decay are 
central, which is mostly the case,  then the forward jet is the $b$-tagged one, however, since a  forward jet 
is less likely to be $b$-tagged, it is clear that the signal is going to be penalized. Unsurprisingly then, the signal corresponding to BP1, BP2 and BP3  survived this constraint at the level of 30.6\%, 33.4\% and 32.2\%, respectively. In spite of the (expected) 
low signal efficiencies, this criterion has been invoked to suppress mainly the four irreducible backgrounds which 
contain more than one $b$-quark in the hard partonic processes: $\nu bbj$, $\nu tb$, $ebbj$ and $ett$. For these channels, 
the survival efficiencies in all the BPs are approximately 24\%, 34\%, 41\% and 34\%, respectively. For $\nu bjj$ and $ebjj$ 
the efficiencies are 14\% and 22\%, respectively. As expected, $\nu jjj$ and $ejjj$ efficiencies were found to  be low,
4.0\% and 5.0\%, respectively, as contributions to signal fakes are here due to a mis-tagged jet. All-in-all, it paid off 
to use this restriction.

\item {\bf Cut E: $N_{\rm jet}$ $\le$ $4$.} After the Cut D stage of cumulative selections, it seems that the  
main backgrounds left are  $\nu tb$, $\nu jjj$, $ebjj$ and $ejjj$. Except $\nu tb$,  the other three  
are due to the huge cross sections to begin with, however, the number of jet distributions show  
that $\nu tb$ has high multiplicity values. To suppress all these backgrounds, we exploited a `number of jet' veto,  
i.e., the event should be removed if it contains more than 4 jets (in practice, we allow one extra  jet only  
compared to the hard process parton multiplicity). The signal events have survival efficiencies  
of 92.2\%, 93.6\% and 95.5\% for BP1, BP2 and BP3, respectively (this shows that, as the charged Higgs mass increases, the  
efficiencies become larger).  The maximal background suppression occurs for events containing a top quark, e.g., 
$\nu tb$ is suppressed by 21.9\%, 16.4\% and 12.1\%  for BP1, BP2 and BP3, respectively. As $ett$  
contains two top quarks,  the suppressions  are more significant, i.e., 70.8\%, 43.1\% and 55.0\%  in correspondence of BP1, BP2 and BP3, 
respectively. The $\nu bbj$, $\nu bjj$, $\nu jjj$($ebbj$, $ebjj$, $ejjj$) channels have a survival rate of 95\%,  93\% and 94\%(96\%, 95\% and  92\%)  for  BP1, BP2 and BP3, respectively. 

\item {\bf Cut F: $N_{\rm{central-jet}}$ $=$ $2$.} After enforcing the above number of jet veto, we see that the $\nu tb$ noise
still produces large contributions,  somewhat higher that  those of the $\nu bbj$ and $\nu bjj$ channels. Notice that, in the signal, out of three jets, 
the two light flavor ones are originating from the charged Higgs boson decays and  
mostly lie in the central regions of the detector while the third one is in the forward region. Now, in comparison, some backgrounds can have 
 a larger 
number of central 
jets, in particular, the $\nu tb$ channel. Here, in fact,  the top quark hadronic decays  would lead to 3 or more jets 
 in the central detector region. Thus, demanding that in the central region we expect exactly two jets suppresses   
the $\nu tb$ noise by 55\% in the case of all  three BPs. Clearly, for the $\nu bbj$ ($\nu bjj$) channels, the reduction 
is less drastic, at the level of 49.6\%, 48.3\% and 45.9\%(59.5\%, 57.7\% and 54.8\%) for the three benchmarks  
respectively. The signal survival rate for this 
criterion is approximately 56\% for all three BPs. The $ebbj$, $ebjj$ and $ejjj$ backgrounds survive at the level of, approximately,    29\%, 52\% and 48\%  for BP2, BP2 and BP3, respectively.
At this point, one can see that the main source of backgrounds are  from $\nu tb$, $\nu jjj$ and $ejjj$,  
the latter two primarily because of the huge cross sections. In the next level of the selection, we would then have to 
ensure  that, out of the  three jets in our signature, the two jets coming from the charged Higgs boson decay are efficiently recognized. 

\item {\bf Cut G}:{ $\mhpm -15$ GeV $\le$  $M_{jj}$ $\le$ $\mhpm +5$ GeV.} In the central region, we reconstruct all 
possible combination of light flavor (i.e., non $b$-tagged) di-jet invariant masses, i.e., $M_{jj}$.  Out of all 
possible combinations, we have chosen the one  for which the absolute difference 
$|M_{jj} - \mhpm|$ (for the particular BP under consideration) is minimized. The ensuing distribution is shown 
in the left panel of Fig.~\ref{mbs} for BP1, BP2  and BP3 (from thicker to thinner) together with that of the 
backgrounds. The peaks of all the signal benchmarks always show up to the left side of the actual $\hpm$ masses 
(for the BPs), an effect  mainly due to jet energy smearing. The mass shifts are dependent upon the jet-cone size 
under consideration, i.e.,   the larger the cone size the more the peak moves to the right. The $M_{jj}$ distributions 
in Fig.~\ref{mbs} further show a rapid fall on the higher side. Therefore, we demanded a somewhat asymmetric mass 
window $\mhpm -15$ GeV $\le$  $M_{jj}$ $\le$ $\mhpm +5$ GeV over which to sample the signals, where the values 
of  $\mhpm$ are set according to the charged Higgs boson mass for the particular BP considered.
\begin{figure}[ht!]
\begin{center}
\raisebox{0.0cm}{\hbox{\includegraphics[angle=0,scale=0.42]{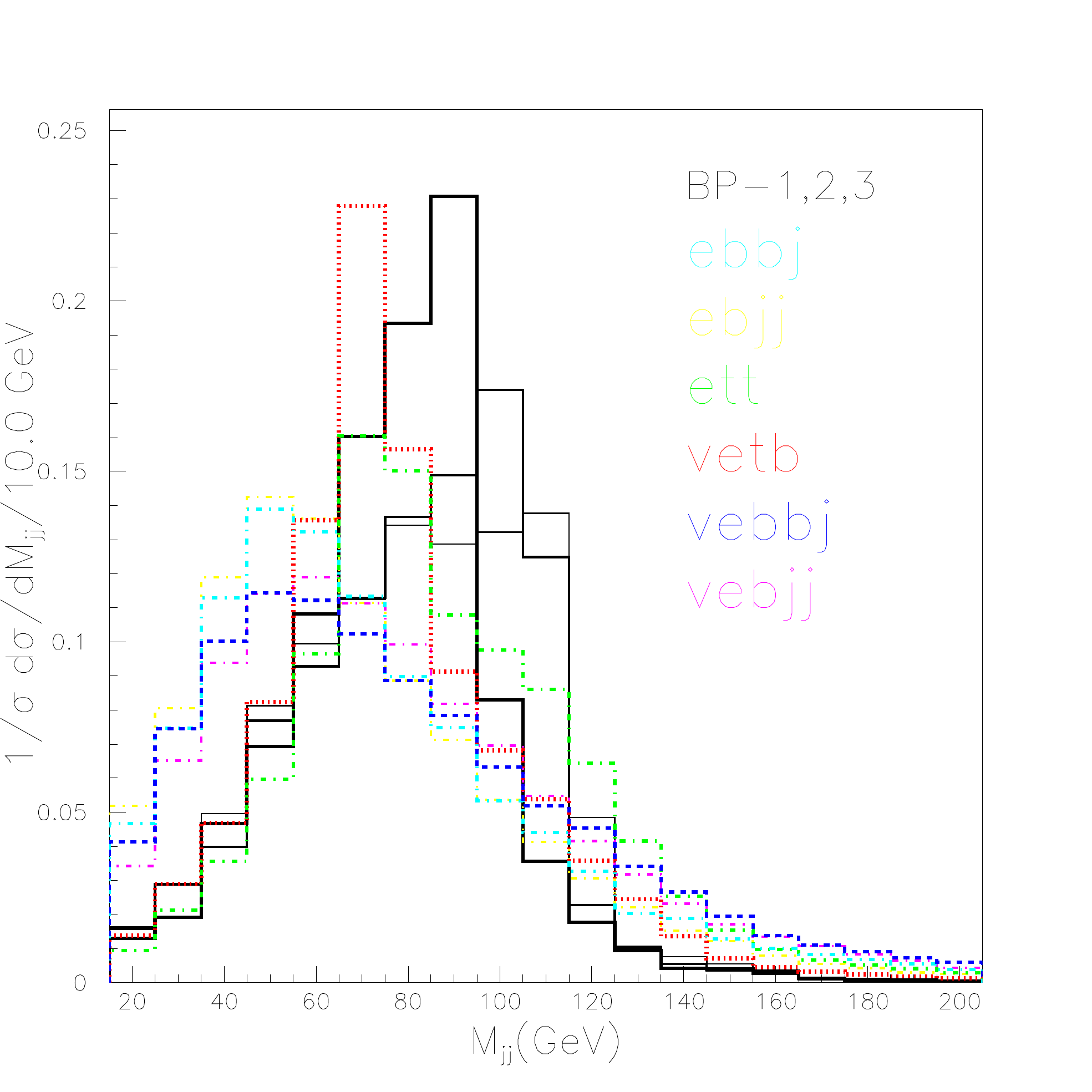}}}
\raisebox{0.0cm}{\hbox{\includegraphics[angle=0,scale=0.42]{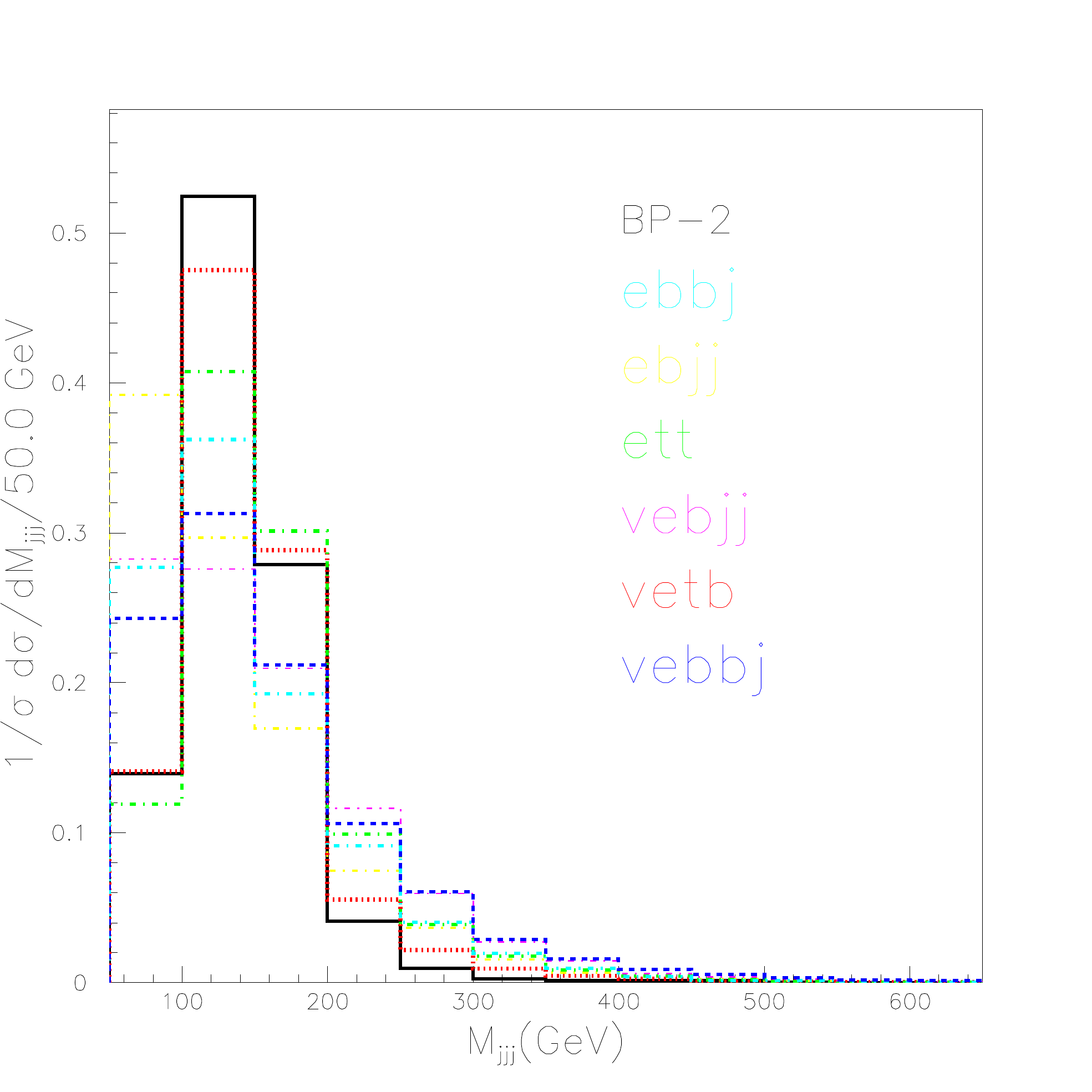}}}
\caption{Left panel: Di-jet invariant mass ($M_{jj}$) distribution for BP1, BP2 and BP3 (from thicker to thinner, i.e., for $\mhpm=$ 98.4, 114.6 and 
121.3 GeV, respectively) as well as the SM backgrounds.  
Right panel: Three-jet invariant mass ($M_{jjj}$) distribution (see the text for the exact definition) for BP2 (it is very similar 
for BP1 and BP3) as well as the SM backgrounds.  For both observables, 
the distributions for $\nu jjj$ ($ejjj$) is very similar to that for $\nu bjj$($ebjj$).  }
\label{mbs}
\end{center}
\end{figure}
The signal efficiencies are approximately 34.8\%, 36.8\% and 33.0\% for BP1, BP2 and 
BP3, respectively. The  $ebbj$ process has a $Z$ boson exchange 
resonant diagram with $Z \to b\bar b$, which shows a mass peak around 60 GeV (approximately 30 GeV 
less than $M_Z$ due to jet energy smearing). The $\nu bjj$ mass peak is somewhat similar 
to the $ebjj$, one as this process has both  $W^\pm$ and $Z$ boson exchange through resonant diagrams. 
Hence, the efficiencies are 13.3\%  for both $\nu bjj$ and $ebjj$, for BP1. For higher values 
of the charged Higgs masses this window selection shifts to higher $M_{jj}$ values, hence the efficiencies 
will be even smaller, e.g.,  for BP2 these are 10.0\% and 11.0\% for $\nu bjj$ and $ebjj$, respectively. A further 1\% is lost by both processes in the case of  BP3.  
The mass reconstruction criterion also suppresses $\nu tb$ to a large extent, i.e.,  approximately 11.7\%, 7.5\% and 7.7\%,  
respectively, for BP1, BP2  and BP3. For the $ejjj$ noise the suppressions are also large, from percent level 
in the case of BP1 and BP2 down to, for BP3, essentially zero. 

\item {\bf Cut H:} {$H_T$ $\ge$ 100 GeV.} We then introduced another  selection based 
on the sum of the transverse momentum of all jets present in the event, $H_{T}$ = $\sum |p_T^j|$. 
The distribution is shown in the left panel of Fig. \ref{hthtv}. The signals show a  peak 
around 125 GeV. The $\nu t b$ noise shows a peak around 135 GeV whereas $e t t$ displays it around 250 GeV (here 
the higher value simply reflects the presence of a larger number of jets). We demanded as selection  $H_T$ $>$ 100 GeV. 
The number of signal events for all the benchmarks remains at approximately 98.0\% or more (for heavy 
Higgs boson masses the survival  probabilities are slightly larger). We see that this selection is not 
reducing much the  $e t t$ channel though, at most by 10.0\%. However, this background is not big at this stage. 
The $ebjj$($ebbj$) background contribution is also not large, so it is not worrisome that it survived our cut at a rate of about  83.0(90.0)\%, e.g.,  for BP2 (numbers are similar for BP1 and BP3). 
At this stage, the dominant background contributions stem from $\nu t b$, $ebjj$, $ejjj$ and $\nu jjj$.
\begin{figure}[ht!]
\begin{center}
\raisebox{0.0cm}{\hbox{\includegraphics[angle=0,scale=0.42]{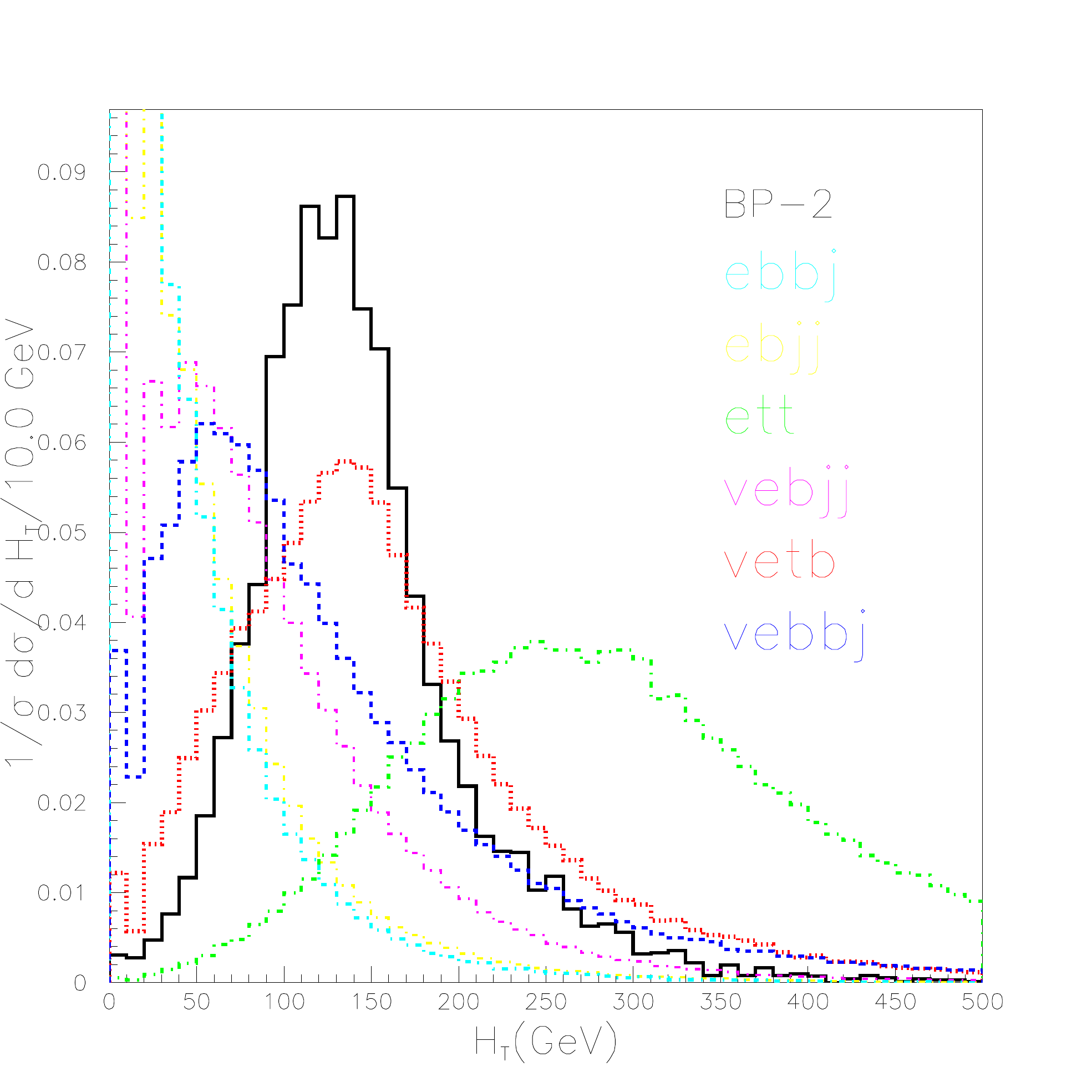}}}
\raisebox{0.0cm}{\hbox{\includegraphics[angle=0,scale=0.42]{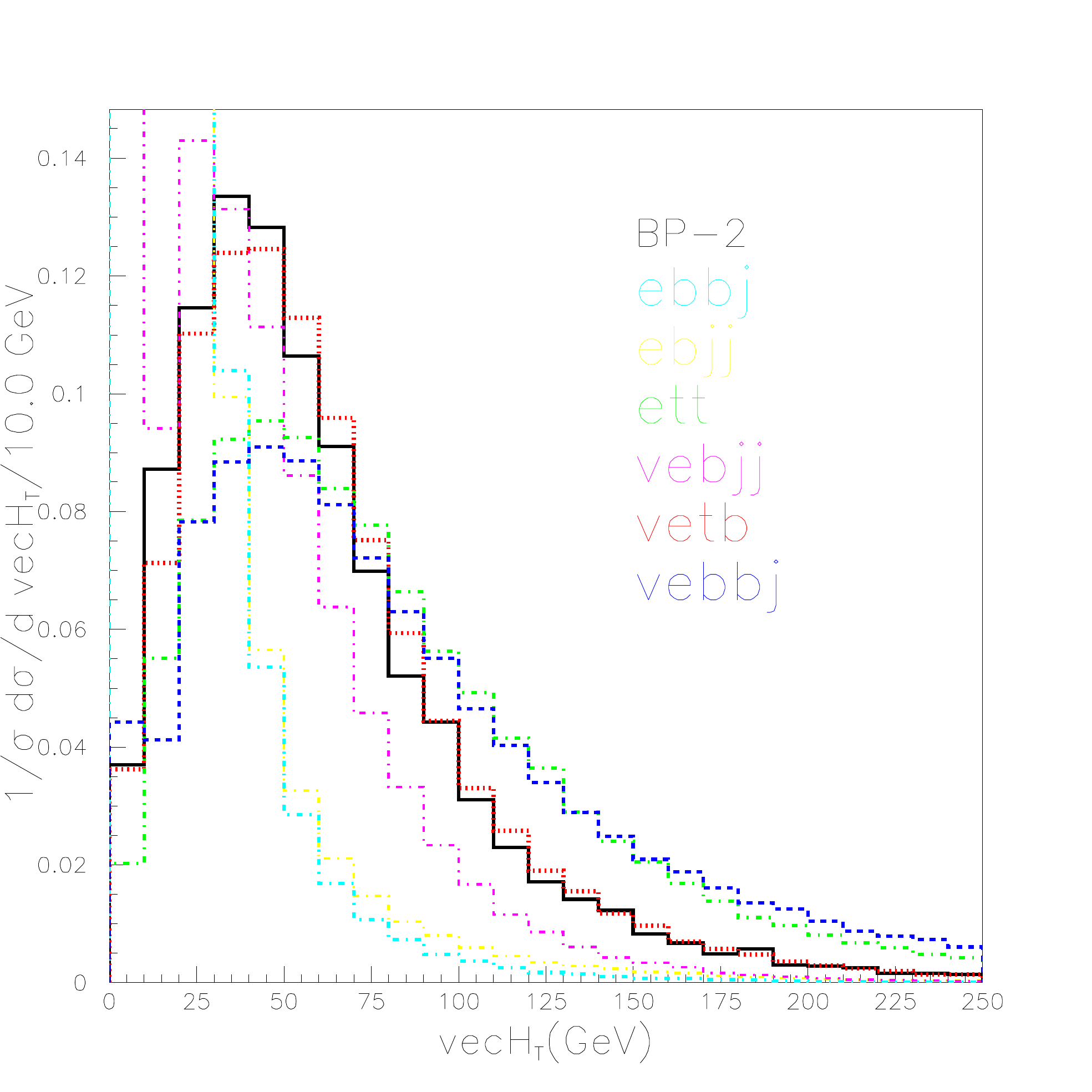}}}
\caption{Left panel: Scalar sum of the jet transverse momentum ($H_T$ = $\sum |p_T^j|$) distribution for BP2 (it is very similar for BP1 and BP3) as well as the SM backgrounds.   Right panel: Vector sum of the jet transverse momentum ($\vec H_T$ = $|\sum \vec p_T^j|$) distribution for BP2   (it is very similar for BP1 and BP3) as well as the SM backgrounds (in the figure the magnitude of the vector is naturally implied).
The distribution for $\nu jjj$($ejjj$) is very similar to that for $\nu bjj$($ebjj$).}
\label{hthtv}
\end{center}
\end{figure}
\item {\bf Cut I:} {100.0 GeV $\le$ $M_{jjj}$ $\le$ 250.0 GeV.} We finally introduce a $3j$  
invariant mass cut, by using as baseline the identified di-jet candidates from the charged Higgs boson decays. With the remaining jets present in the event, we 
constructed all possible combinations, chosen the minimum one and shown it in the right panel of  Fig.~\ref{mbs}. 
All the signal BPs survive this cut at the rate of  approximately 99.0\% while a somewhat more significant reduction take  
places for $ebbj$ and  $\nu jjj$,  77.8\% and 93.1\%, respectively, e.g., for 
 BP2. The trend is similar for BP1 and BP3.  

\end{itemize}
\begin{table}[t!]
\centering
{\tiny
\begin{tabular}{||c||c|c||c|c|c|c|c|c|c|c|c|c||}
\hline
\hline
Proc,$\mhpm$&EvtSim&EvtRaw&A&B&C&D&E&F&G&H&I&${\cal S}$\\
\hline
\hline
BP1,98.4& 100K& 1098.8 & 770.0& 717.7& 628.0& 192.0&   176.9&   105.1&  36.6&   35.9&   35.7&0.33(1.06)\\
\hline
$\nu bbj$&200K&53700.0 & 14181.3&     13377.1&13054.1&      3276.5&  3079.9&  1527.8&   198.2&  179.1&  165.9&\\
$\nu bjj$&250K&49000.0&15069.1& 14687.2& 13370.5& 1912.4&   1747.4& 1039.7&  138.1&   119.5&   112.3&\\
$\nu t b$&250K&814000.0& 547296.7&    461566.3& 412151.4&    141081.6&110126.8& 49234.7&  4383.9& 4221.1& 4093.2&\\ 
$\nu jjj$&250K&13737000.0& 3773162.4&   3743307.5& 3541741.3&    131141.7&118870.0& 52658.1&  6135.8& 5311.6& 4945.3&11277.9(106.2)\\
$e bbj$& 250K&10613000.0 & 1079324.1&    211548.9& 78829.7&     33607.3& 31484.7&  9433.6&  1179.2& 1061.3& 1061.3&\\
$e bjj$&250K&1566000.0&  197721.6&     43283.9&     12809.8&      2662.2&  2530.6&  1472.0&   194.2&  150.3&  150.3& \\
$e t \bar t$&200K& 39500.0& 37140.8&  3120.9& 2267.2&777.4& 227.1& 79.9& 12.6& 11.2&   10.3& \\
$e jjj$&250K&133081000.0 & 16971980.7& 4144664.7&   1148913.5&56928.1& 51752.9& 17743.8&1478.7&739.3&739.3&\\
\hline
\hline
BP2,114.6& 100K&3200.5&1868.4& 1786.7& 1560.1& 521.2& 487.9& 271.5&100.0&99.1&   98.2&1.01(3.21)\\
\hline
$\nu bbj$&200K&53700.0& 10831.5& 10166.2&      9938.5&      2417.0&  2313.9& 1117.8&   119.8&  112.0&  105.3&\\
$\nu bjj$&250K&49000.0&  11131.1&      10817.2&       9980.3&  1372.4& 1279.1& 737.5&     74.1&    68.3&    62.1& \\
$\nu t b$&250K&814000.0& 481904.4&    409285.0&    367666.9&    124999.5&104504.5& 46281.1&  3471.1& 3325.7& 3087.3&\\
$\nu jjj$&250K&13737000.0& 2741978.5&   2718442.6&   2596001.0&     94235.2& 88557.3& 37822.3&  4029.5& 3663.2& 3388.4&9376.3(96.8)\\
$e bbj$& 250K&10613000.0 &697970.0&    126115.7&     54774.0& 21933.2& 21107.7&  6072.9&   589.6&  530.6&  412.7&\\
$e bjj$&250K&1566000.0&126694.7& 24592.3& 8919.9& 2054.6&  1960.6&  1027.3&   112.8&   94.0&   87.7& \\ 
$e t \bar t$&200K& 39500.0& 36197.8& 3060.7& 2302.6& 787.3&290.9&   108.2&    18.0&   16.2&   14.8&\\
$e jjj$&250K&133081000.0 &10292904.5& 2348101.2&    825088.4& 36966.3& 34009.0& 16265.2&  2957.3& 2218.0& 2218.0&\\
\hline
\hline
BP3,121.3& 100K& 1900.5& 848.4&  803.4&  710.1& 228.4&   218.2&  120.2&    39.6&   39.6& 39.4&0.55(1.75)\\
\hline
$\nu bbj$&200K&53700.0&8002.9& 7472.6&  7323.8& 1718.9&  1657.4& 761.2&    82.7&   78.7&  70.9&\\
$\nu bjj$&250K&49000.0& 7885.2&7638.1&  7126.3& 936.2&    889.1&  487.4&     54.3&    51.2&    45.2& \\
$\nu t b$&250K&814000.0& 400935.8& 341811.2& 309728.4& 104364.9& 91742.3& 40519.2&  3133.9& 2953.6& 2657.1 &\\
$\nu jjj$&250K&13737000.0& 1937728.3& 1919046.1& 1848896.3& 61724.5& 58519.2& 22803.3&  2197.9& 2014.7& 1740.0&5058.7(71.1)\\
$e bbj$& 250K&10613000.0 & 438014.8&  74289.8& 37027.0& 15801.3& 15093.8& 4422.0&   471.7&  471.7&  471.7&\\
$e bjj$&250K&1566000.0&79571.0&14050.0& 6107.4& 1434.4&  1384.3&   670.2& 81.4&   68.9&   62.6(0.909)&\\
$e t \bar t$&200K& 39500.0&34770.0& 2989.8& 2348.0&779.2&   351.5&   131.5&    14.8&   14.4&   11.2&\\ 
$e jjj$&250K&133081000.0 & 6169680.2&   1316740.6& 575935.4&     37705.7& 36966.3& 11829.2&   0.0& 0.0& 0.0&\\
\hline
\hline
\end{tabular}
}
\caption{Cut flow for $S$ and $B$ events  in the $3j + \met$ channel  for BP1, BP2 and BP3 as well as the SM backgrounds alongside the final significance ${\cal S}$ for 100 fb$^{-1}$(1 ab$^{-1}$) of luminosity. The numbers in the final column in the SM background block 
represent the total contributions from all the backgrounds ($B$) for 100 fb$^{-1}$ while in the parentheses we present  $\sqrt{B}$.
The entries for $\mhpm$ in the first column are in GeV.}
\label{tab:sigb}
\end{table}

It is rather clear from Tab. \ref{tab:sigb} that the significances ${\cal S}$ in the three-jet with missing energy channel 
are approximately 0.33, 1.01 and 0.55  at the FCC-eh with 100 fb$^{-1}$ luminosity, in correspondence of  BP1, BP2 and BP3, 
respectively, following the cut based cumulative selections that we have just described. With 1 ab$^{-1}$, the significances would be 1.1, 3.2 and 1.8, for a charged Higgs mass of 98.4, 114.6 and 121.3 GeV, respectively (as in BP1, BP2 and BP3). In short, these are not very promising for the FCC-eh in standard configurations. Clearly, any machine improvement leading to higher luminosity will prove useful. However, this cannot be counted upon at this stage.  

\begin{figure}[ht!]
\begin{center}
\raisebox{0.0cm}{\hbox{\includegraphics[angle=0,scale=0.42]{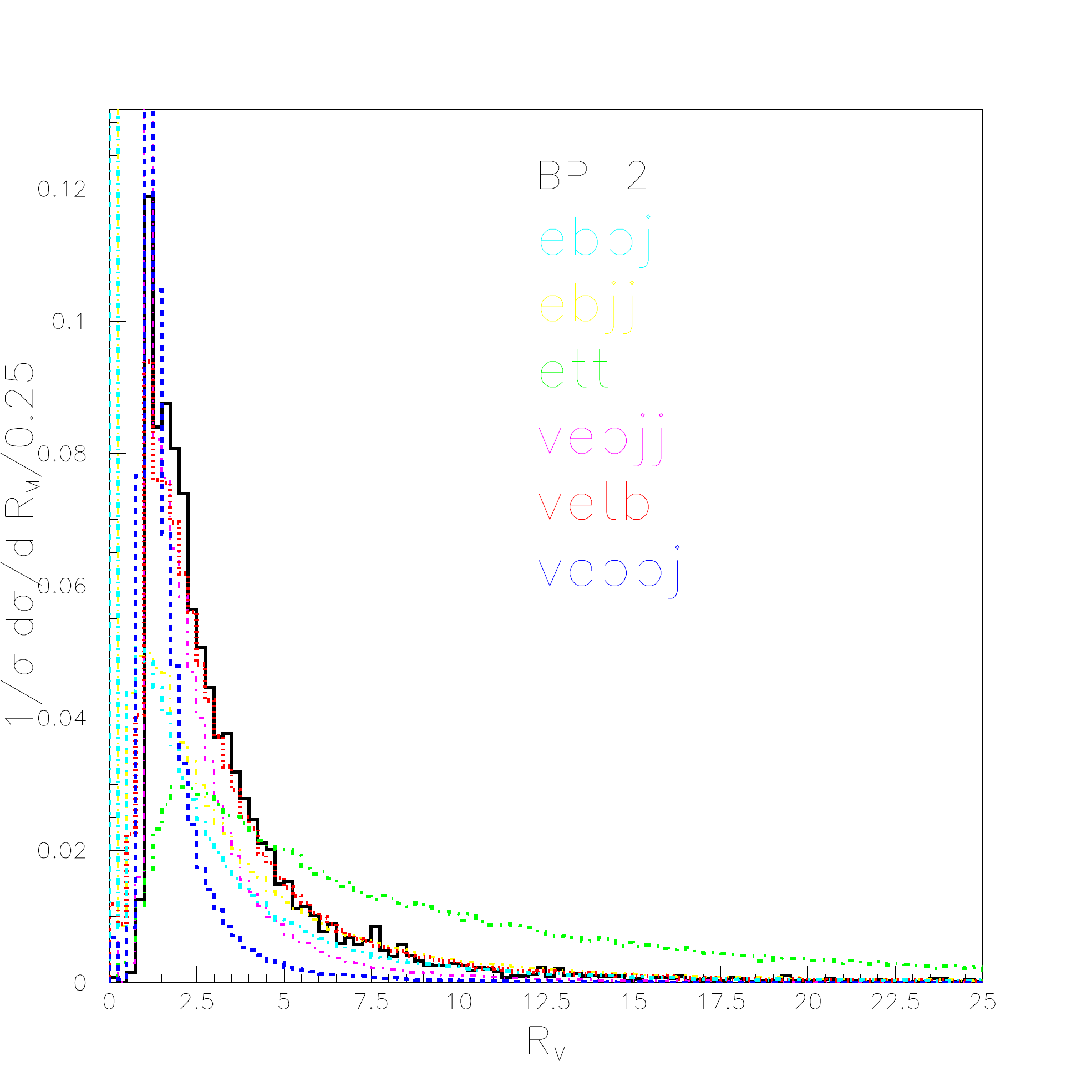}}}
\raisebox{0.0cm}{\hbox{\includegraphics[angle=0,scale=0.42]{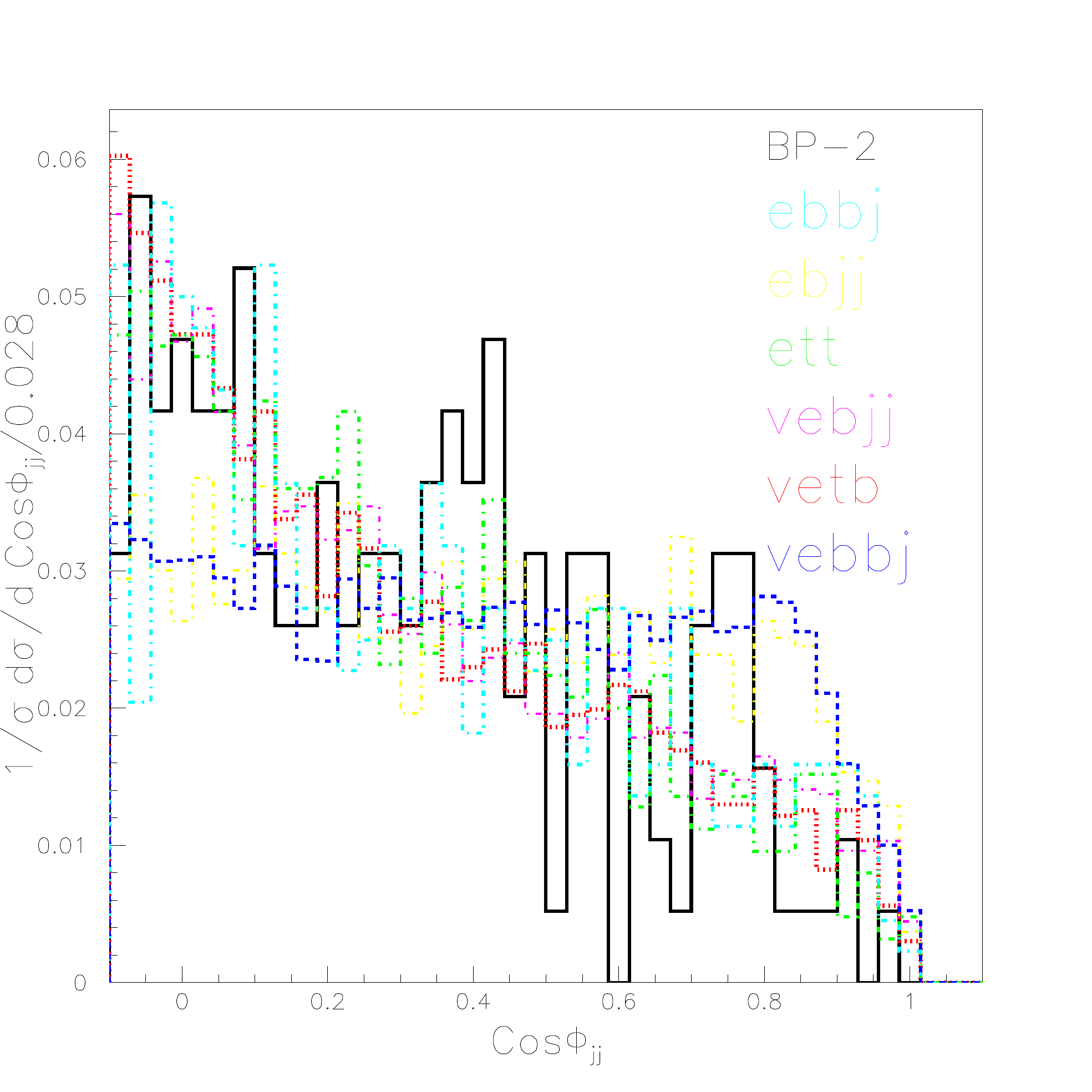}}}
\caption{Left panel: $R_M$ (=  $H_T$/$\met$) distribution for BP2  (it is very similar for BP1 and BP3) as well as the SM backgrounds. 
Right panel: cos$\phi_{jj}$ (see the text) distribution for BP2  (it is very similar for BP1 and BP3) as well as for the SM backgrounds. 
The distribution for $\nu jjj$($ejjj$) is very similar to that for $\nu bjj$($ebjj$).
}
\label{rmcosphi}
\end{center}
\end{figure}

Hence, to find better significances, we have to depart from this approach (based on previous similar work of ours) and 
exploit an optimization technique, as follows. First of all, we keep the selection 
as it is up to cut F, i.e., up-to the requirement of the presence of exactly two central jets. Afterwards, we vary the 
following kinematical variables: $\met$ (the missing transverse energy), $H_T$ (the scalar sum of the transverse momenta of the jets), $|\vec H_T|$ 
(the vector sum of the transverse momenta of the jets), $M_{jj}$ (i.e., the di-jet invariant mass of the charged Higgs boson candidates), 
$R_M$ = $H_T$/$\met$ plus, among all possible combinations of jets present in an event, the minimum of cos($\phi_{jj}$) (i.e., the cosine of  
the azimuthal angle between jets) and the corresponding  $\Delta R$($\eta_{jj},\phi_{jj}$). The distributions of 
$R_M$ and cos($\phi_{jj}$) are shown in the left and right panel of Fig.~\ref{rmcosphi}, respectively.

The numerical values of all these kinematical variables are then varied over a multi-dimensional grid each within a minimum and maximum range (determined by investigating  
the corresponding distributions). In particular, we adopted the following kinematical ranges: $\met$ in $(20.0,40.0)$ GeV 
with step-size 5.0 GeV; $H_T$ in $(95.0,110.0)$ GeV  with step-size 5.0 GeV; $|\vec H_T|$ in $(20.0,40.0)$ GeV with step-size 5.0 GeV; 
$R_M$ in $(2.5,3.5)$ with step-size 0.025; the upper value (end point) of the $M_{jj}$ spectrum in ($\mhpm$, $\mhpm$+10.0 GeV) 
with step-size 2.5 GeV; the lower value of the $M_{jj}$ spectrum  in ($\mhpm$--25.0 GeV, $\mhpm$--15.0 GeV) with step-size 2.5 GeV; 
the upper value of cos($\phi_{jj}$) in $(0.45,0.55)$ with step-size 0.01 and the upper value of $\Delta R$($\eta_{jj}, \phi_{jj}$) in 
$(2.1,3.5)$ with step-size 0.1. For each of the generated combinations of such variables, each used as new kinematical constraint from Cut G onwards, also
accompanied by additional cuts in
 $R_M$, $\cos(\phi_{jj})$ and $\Delta R$($\eta_{jj},\phi_{jj}$), we then estimated 
the number of signal events, of  total  background ones plus finally the significances ${\cal S}$. As shown in Tab.~\ref{tab:optlep}, there indeed 
exist combinations for which both evidence and (near) discovery of our $\hpm$ signals can be established, albeit only at 1 ab$^{-1}$ of luminosity. Notice that this
optimization is rather robust, as it is not biased by the acceptances (in pseudo-rapidity and transverse momentum) or previous selection Cuts A--F, as we have verified explicitly.

\begin{table}[t!]
\centering
\scalebox{0.7}{
\begin{tabular}{||c||c|c|c|c|c|c|c|c|c|c|c||}
\hline
BP, $\mhpm$&$\met$&$H_T$&$\vec H_T$&$M_{jj}$$<$&$M_{jj}$$>$&$R_M$$<$& cos($\phi_{jj}$)$<$ & $\Delta R$($\phi_{jj}$)$<$&S&B&${\cal S}$\\
\hline
\hline
& 20.00 &105.00 & 20.00 &98.40 & 80.90 &  2.50 &  0.52 &  2.10 & 35.1& 9055.5 & 0.37(1.18)\\
BP1, 98.4&  20.00 &100.00 & 20.00 & 98.40 & 75.90 & 2.50 &  0.52 &  3.10 & 49.4 & 19714.8 & 0.35 (1.12)\\
& 20.00 &105.00 & 20.00 & 103.40 & 73.40 &  2.50 &  0.52 &  3.10 & 55.8 & 27072.3 & 0.34(1.08)\\
\hline
\hline
& 20.00 &110.00 & 20.00 &114.60 & 89.60 &  2.50 &  0.54 &  2.90&145.3 & 11027.8 & 1.38(4.43)\\
BP2, 114.6& 20.00 &110.00 & 20.00 &114.60 & 99.60 &  2.50 &  0.54 &  2.20 & 86.6& 4890.5&1.24(3.96) \\
& 30.00 &110.00 & 30.00 &114.60 & 97.10 &  2.50 &  0.45 &  2.10 & 74.7 & 5005.8 & 1.05(3.38)\\
\hline
\hline
& 20.00 & 95.00 & 20.00 & 121.30 & 96.30 &  2.50 &  0.45 & 2.80 & 61.5& 8327.2 & 0.67(2.16)\\
BP3, 121.3& 20.00 &110.00 & 20.00 & 121.30 & 96.30  &  2.50 & 0.45 &  2.20 & 54.7& 7040.8  & 0.65(2.08)\\
&20.00 &100.00 & 20.00 & 121.30 &103.80 &  2.50 &  0.45 &  2.60 & 44.4& 5234.5& 0.61(1.96)\\
\hline
\hline
\end{tabular}
}
\caption{The best three optimization configurations for BP1, BP2 and BP3 signals  for different  
sets of kinematical selection cuts in $\met$, $H_T$, $\vec H_T$, $M_{jj}$, $R_M$,  cos($\phi_{jj}$) and  
$\Delta R$($\eta_{jj},\phi_{jj}$) (see the text for details) with the best significances obtained for 100 fb$^{-1}$(1 ab$^{-1}$).
All numerical entries are in GeV except those in the last six columns.
}
\label{tab:optlep}
\end{table}

\section{Conclusions}
\label{sec:conclusion}

Among the experimental facilities which might soon become operational and potentially 
competitive to look for  any kind of Higgs bosons, be it neutral and/or charged,  the FCC-eh machine 
(via $e^- p$ collisions) to be  located at CERN is one that is presently receiving significant attention from the particle physics community. 
In this work, we have considered the NMSSM  with low mass charged Higgs bosons, $\hpm$, wherein the second lightest neutral CP-even 
Higgs boson, $h_2$, can be identified with the SM-like Higgs boson discovered at CERN in 2012. In the NMSSM, this is  naturally possible in presence of  all most accurate theoretical and 
most up-to-date experimental constraints, the latter from both lower energy experiments as well as recent LHC data about Supersymmetric  
and Higgs state searches by ATLAS and CMS, indeed combined with precision measurements of the SM-like Higgs boson properties by the same experimental collaborations. The model also has naturally a low mass lightest neutralino 
($\lsp$), which serves as the possible candidate for cold DM, so that we have further  imposed experimental  limits coming from  WMAP, Planck and (in)direct DM searches.

We have in particular studied the $e^-p\to \nu_e h^- b$ production mechanism, followed by the $h^- \to s\bar u + s\bar c$  decay modes
at the FCC-eh. In our analysis, we have  first performed a detailed NMSSM
parameter space scan by using {\tt NMSSMTools}. For the allowed parameters we have then 
estimated the $h^-$ production cross section (folded with the BR) to find 
the  inclusive signal  rates at the FCC-eh facility. These were found substantial, amongst competing $\hpm$ decay modes, so that a signal-to-background analysis was envisaged.  The $h^-$ signature selected for consideration was $3j + \met$. Herein, two jets  originate 
from the charged Higgs boson decay, $h^- \to  s\bar u + s\bar c$ The remaining jet originates from 
the remnant of the proton flux which is likely be at  large rapidity (in the forward 
or backward region, but not both). This could be a light flavor jet but it would mostly be a $b$-jet. (Hence we have eventually demanded exactly one  
$b$-tagged jet, including proper mis-tagging rates.) We have also  considered the reducible and irreducible SM 
backgrounds stemming from both  charged- 
($\nu tb$, $\nu bb j$, $\nu bjj$ and $\nu jjj$) and neutral-current 
($ebbj$, $ebjj$, $ett$ and $ejjj$) processes.

We  have then performed a full  Monte Carlo  simulation using {\tt MadGraph} at the parton level 
followed by {\tt PYTHIA} as the parton shower/hadronization/heavy flavor decay event generator of choice and its 
{\tt PYCELL} toy calorimeter simulation modified in accordance with the FCC-eh (similar to the LHeC) detector 
parameters. Upon defining a selection procedure importing elements from previous work of ours for the LHeC, we have found interesting results. 
The latter in fact   show that,  with 1 ab$^{-1}$ of luminosity,  charged Higgs bosons with mass close to 114(121) GeV can be extracted with  
 maximal significances of 3.2(1.8)$\sigma$ using normal cut based selections. To enhance 
these significances, though, we have finally adopted an optimization technique exploiting all the  kinematical observables devised for the (sequential) cut based selection in a correlated approach to find that for  the above  charged Higgs masses 
of 114(121) GeV the significances can go up to 4.4(2.2)$\sigma$. In contrast, $\hpm$ masses closer to $M_{W^\pm}$, e.g. 98 GeV, remain unaccessible.

If the FCC-eh experimental collaborations would invoke more complex discriminators and/or use multi-variate analysis, 
one can naturally expect a discovery with 5$\sigma$ for low mass charged Higgs bosons  in the regions of 115 GeV with 
1 ab$^{-1}$ of integrated luminosity in $h^-$ decays which could be attributed to an underlying NMSSM construct.

\subsubsection*{Acknowledgments}  

SPD acknowledges the High Performance Computing (HPC) facility at Universidad de los Andes. SM is supported in 
part by the NExT Institute, STFC Consolidated Grant ST/L000296/1 and H2020-MSCA-RISE-2014 grant no. 645722  
(NonMinimalHiggs). AR and JHS are supported by SNI-CONACYT (M\'exico), VIEP-BUAP and by PRODEP-SEP (M\'exico)  
under the grant: ``Red Tem\'atica: F\'{\i}sica del Higgs y del Sabor".

\bibliography{hpmB.bib} 

\begin{thebibliography}{56}%
\makeatletter
\providecommand \@ifxundefined [1]{%
 \@ifx{#1\undefined}
}%
\providecommand \@ifnum [1]{%
 \ifnum #1\expandafter \@firstoftwo
 \else \expandafter \@secondoftwo
 \fi
}%
\providecommand \@ifx [1]{%
 \ifx #1\expandafter \@firstoftwo
 \else \expandafter \@secondoftwo
 \fi
}%
\providecommand \natexlab [1]{#1}%
\providecommand \enquote  [1]{``#1''}%
\providecommand \bibnamefont  [1]{#1}%
\providecommand \bibfnamefont [1]{#1}%
\providecommand \citenamefont [1]{#1}%
\providecommand \href@noop [0]{\@secondoftwo}%
\providecommand \href [0]{\begingroup \@sanitize@url \@href}%
\providecommand \@href[1]{\@@startlink{#1}\@@href}%
\providecommand \@@href[1]{\endgroup#1\@@endlink}%
\providecommand \@sanitize@url [0]{\catcode `\\12\catcode `\$12\catcode
  `\&12\catcode `\#12\catcode `\^12\catcode `\_12\catcode `\%12\relax}%
\providecommand \@@startlink[1]{}%
\providecommand \@@endlink[0]{}%
\providecommand \url  [0]{\begingroup\@sanitize@url \@url }%
\providecommand \@url [1]{\endgroup\@href {#1}{\urlprefix }}%
\providecommand \urlprefix  [0]{URL }%
\providecommand \Eprint [0]{\href }%
\providecommand \doibase [0]{http://dx.doi.org/}%
\providecommand \selectlanguage [0]{\@gobble}%
\providecommand \bibinfo  [0]{\@secondoftwo}%
\providecommand \bibfield  [0]{\@secondoftwo}%
\providecommand \translation [1]{[#1]}%
\providecommand \BibitemOpen [0]{}%
\providecommand \bibitemStop [0]{}%
\providecommand \bibitemNoStop [0]{.\EOS\space}%
\providecommand \EOS [0]{\spacefactor3000\relax}%
\providecommand \BibitemShut  [1]{\csname bibitem#1\endcsname}%
\let\auto@bib@innerbib\@empty
\bibitem [{\citenamefont {Aad}\ \emph {et~al.}(2012)\citenamefont {Aad} \emph
  {et~al.}}]{Aad:2012tfa}%
  \BibitemOpen
  \bibfield  {author} {\bibinfo {author} {\bibfnamefont {G.}~\bibnamefont
  {Aad}} \emph {et~al.} (\bibinfo {collaboration} {ATLAS}),\ }\href {\doibase
  10.1016/j.physletb.2012.08.020} {\bibfield  {journal} {\bibinfo  {journal}
  {Phys. Lett.}\ }\textbf {\bibinfo {volume} {B716}},\ \bibinfo {pages} {1}
  (\bibinfo {year} {2012})},\ \Eprint {http://arxiv.org/abs/1207.7214}
  {arXiv:1207.7214 [hep-ex]} \BibitemShut {NoStop}%
\bibitem [{\citenamefont {Chatrchyan}\ \emph {et~al.}(2012)\citenamefont
  {Chatrchyan} \emph {et~al.}}]{Chatrchyan:2012xdj}%
  \BibitemOpen
  \bibfield  {author} {\bibinfo {author} {\bibfnamefont {S.}~\bibnamefont
  {Chatrchyan}} \emph {et~al.} (\bibinfo {collaboration} {CMS}),\ }\href
  {\doibase 10.1016/j.physletb.2012.08.021} {\bibfield  {journal} {\bibinfo
  {journal} {Phys. Lett.}\ }\textbf {\bibinfo {volume} {B716}},\ \bibinfo
  {pages} {30} (\bibinfo {year} {2012})},\ \Eprint
  {http://arxiv.org/abs/1207.7235} {arXiv:1207.7235 [hep-ex]} \BibitemShut
  {NoStop}%
\bibitem [{\citenamefont {Aad}\ \emph {et~al.}(2016{\natexlab{a}})\citenamefont
  {Aad} \emph {et~al.}}]{Khachatryan:2016vau}%
  \BibitemOpen
  \bibfield  {author} {\bibinfo {author} {\bibfnamefont {G.}~\bibnamefont
  {Aad}} \emph {et~al.} (\bibinfo {collaboration} {ATLAS, CMS}),\ }\href
  {\doibase 10.1007/JHEP08(2016)045} {\bibfield  {journal} {\bibinfo  {journal}
  {JHEP}\ }\textbf {\bibinfo {volume} {08}},\ \bibinfo {pages} {045} (\bibinfo
  {year} {2016}{\natexlab{a}})},\ \Eprint {http://arxiv.org/abs/1606.02266}
  {arXiv:1606.02266 [hep-ex]} \BibitemShut {NoStop}%
\bibitem [{\citenamefont {Ellis}\ and\ \citenamefont
  {You}(2013)}]{Ellis:2013lra}%
  \BibitemOpen
  \bibfield  {author} {\bibinfo {author} {\bibfnamefont {J.}~\bibnamefont
  {Ellis}}\ and\ \bibinfo {author} {\bibfnamefont {T.}~\bibnamefont {You}},\
  }\href {\doibase 10.1007/JHEP06(2013)103} {\bibfield  {journal} {\bibinfo
  {journal} {JHEP}\ }\textbf {\bibinfo {volume} {06}},\ \bibinfo {pages} {103}
  (\bibinfo {year} {2013})},\ \Eprint {http://arxiv.org/abs/1303.3879}
  {arXiv:1303.3879 [hep-ph]} \BibitemShut {NoStop}%
\bibitem [{\citenamefont {de~Blas}\ \emph {et~al.}(2016)\citenamefont
  {de~Blas}, \citenamefont {Ciuchini}, \citenamefont {Franco}, \citenamefont
  {Mishima}, \citenamefont {Pierini}, \citenamefont {Reina},\ and\
  \citenamefont {Silvestrini}}]{deBlas:2016ojx}%
  \BibitemOpen
  \bibfield  {author} {\bibinfo {author} {\bibfnamefont {J.}~\bibnamefont
  {de~Blas}}, \bibinfo {author} {\bibfnamefont {M.}~\bibnamefont {Ciuchini}},
  \bibinfo {author} {\bibfnamefont {E.}~\bibnamefont {Franco}}, \bibinfo
  {author} {\bibfnamefont {S.}~\bibnamefont {Mishima}}, \bibinfo {author}
  {\bibfnamefont {M.}~\bibnamefont {Pierini}}, \bibinfo {author} {\bibfnamefont
  {L.}~\bibnamefont {Reina}}, \ and\ \bibinfo {author} {\bibfnamefont
  {L.}~\bibnamefont {Silvestrini}},\ }\href {\doibase 10.1007/JHEP12(2016)135}
  {\bibfield  {journal} {\bibinfo  {journal} {JHEP}\ }\textbf {\bibinfo
  {volume} {12}},\ \bibinfo {pages} {135} (\bibinfo {year} {2016})},\ \Eprint
  {http://arxiv.org/abs/1608.01509} {arXiv:1608.01509 [hep-ph]} \BibitemShut
  {NoStop}%
\bibitem [{\citenamefont {Kane}\ \emph {et~al.}(1994)\citenamefont {Kane},
  \citenamefont {Kolda}, \citenamefont {Roszkowski},\ and\ \citenamefont
  {Wells}}]{Kane:1993td}%
  \BibitemOpen
  \bibfield  {author} {\bibinfo {author} {\bibfnamefont {G.~L.}\ \bibnamefont
  {Kane}}, \bibinfo {author} {\bibfnamefont {C.~F.}\ \bibnamefont {Kolda}},
  \bibinfo {author} {\bibfnamefont {L.}~\bibnamefont {Roszkowski}}, \ and\
  \bibinfo {author} {\bibfnamefont {J.~D.}\ \bibnamefont {Wells}},\ }\href
  {\doibase 10.1103/PhysRevD.49.6173} {\bibfield  {journal} {\bibinfo
  {journal} {Phys. Rev.}\ }\textbf {\bibinfo {volume} {D49}},\ \bibinfo {pages}
  {6173} (\bibinfo {year} {1994})},\ \Eprint
  {http://arxiv.org/abs/hep-ph/9312272} {arXiv:hep-ph/9312272 [hep-ph]}
  \BibitemShut {NoStop}%
\bibitem [{\citenamefont {Gunion}\ \emph {et~al.}(2000)\citenamefont {Gunion},
  \citenamefont {Haber}, \citenamefont {Kane},\ and\ \citenamefont
  {Dawson}}]{Gunion:1989we}%
  \BibitemOpen
  \bibfield  {author} {\bibinfo {author} {\bibfnamefont {J.~F.}\ \bibnamefont
  {Gunion}}, \bibinfo {author} {\bibfnamefont {H.~E.}\ \bibnamefont {Haber}},
  \bibinfo {author} {\bibfnamefont {G.~L.}\ \bibnamefont {Kane}}, \ and\
  \bibinfo {author} {\bibfnamefont {S.}~\bibnamefont {Dawson}},\ }\href@noop {}
  {\bibfield  {journal} {\bibinfo  {journal} {Front. Phys.}\ }\textbf {\bibinfo
  {volume} {80}},\ \bibinfo {pages} {1} (\bibinfo {year} {2000})}\BibitemShut
  {NoStop}%
\bibitem [{\citenamefont {Kim}\ and\ \citenamefont
  {Nilles}(1984)}]{Kim:1983dt}%
  \BibitemOpen
  \bibfield  {author} {\bibinfo {author} {\bibfnamefont {J.~E.}\ \bibnamefont
  {Kim}}\ and\ \bibinfo {author} {\bibfnamefont {H.~P.}\ \bibnamefont
  {Nilles}},\ }\href {\doibase 10.1016/0370-2693(84)91890-2} {\bibfield
  {journal} {\bibinfo  {journal} {Phys. Lett.}\ }\textbf {\bibinfo {volume}
  {138B}},\ \bibinfo {pages} {150} (\bibinfo {year} {1984})}\BibitemShut
  {NoStop}%
\bibitem [{\citenamefont {Ellwanger}\ \emph {et~al.}(2005)\citenamefont
  {Ellwanger}, \citenamefont {Gunion},\ and\ \citenamefont
  {Hugonie}}]{Ellwanger:2004xm}%
  \BibitemOpen
  \bibfield  {author} {\bibinfo {author} {\bibfnamefont {U.}~\bibnamefont
  {Ellwanger}}, \bibinfo {author} {\bibfnamefont {J.~F.}\ \bibnamefont
  {Gunion}}, \ and\ \bibinfo {author} {\bibfnamefont {C.}~\bibnamefont
  {Hugonie}},\ }\href {\doibase 10.1088/1126-6708/2005/02/066} {\bibfield
  {journal} {\bibinfo  {journal} {JHEP}\ }\textbf {\bibinfo {volume} {02}},\
  \bibinfo {pages} {066} (\bibinfo {year} {2005})},\ \Eprint
  {http://arxiv.org/abs/hep-ph/0406215} {arXiv:hep-ph/0406215 [hep-ph]}
  \BibitemShut {NoStop}%
\bibitem [{\citenamefont {Barbieri}\ \emph {et~al.}(2013)\citenamefont
  {Barbieri}, \citenamefont {Buttazzo}, \citenamefont {Kannike}, \citenamefont
  {Sala},\ and\ \citenamefont {Tesi}}]{Barbieri:2013hxa}%
  \BibitemOpen
  \bibfield  {author} {\bibinfo {author} {\bibfnamefont {R.}~\bibnamefont
  {Barbieri}}, \bibinfo {author} {\bibfnamefont {D.}~\bibnamefont {Buttazzo}},
  \bibinfo {author} {\bibfnamefont {K.}~\bibnamefont {Kannike}}, \bibinfo
  {author} {\bibfnamefont {F.}~\bibnamefont {Sala}}, \ and\ \bibinfo {author}
  {\bibfnamefont {A.}~\bibnamefont {Tesi}},\ }\href {\doibase
  10.1103/PhysRevD.87.115018} {\bibfield  {journal} {\bibinfo  {journal} {Phys.
  Rev.}\ }\textbf {\bibinfo {volume} {D87}},\ \bibinfo {pages} {115018}
  (\bibinfo {year} {2013})},\ \Eprint {http://arxiv.org/abs/1304.3670}
  {arXiv:1304.3670 [hep-ph]} \BibitemShut {NoStop}%
\bibitem [{\citenamefont {Miller}\ \emph {et~al.}(2005)\citenamefont {Miller},
  \citenamefont {Moretti},\ and\ \citenamefont {Nevzorov}}]{Miller:2005qua}%
  \BibitemOpen
  \bibfield  {author} {\bibinfo {author} {\bibfnamefont {D.~J.}\ \bibnamefont
  {Miller}}, \bibinfo {author} {\bibfnamefont {S.}~\bibnamefont {Moretti}}, \
  and\ \bibinfo {author} {\bibfnamefont {R.}~\bibnamefont {Nevzorov}},\ }in\
  \href
  {http://theory.sinp.msu.ru/~qfthep04/2004/Proceedings2004/Miller_QFTHEP04_212_219.ps}
  {\emph {\bibinfo {booktitle} {{High energy physics and quantum field theory.
  Proceedings, 18th International Workshop, QFTHEP 1004, St. Petersburg,
  Russia, June 17-23, 2004}}}}\ (\bibinfo {year} {2005})\ pp.\ \bibinfo {pages}
  {212--219},\ \Eprint {http://arxiv.org/abs/hep-ph/0501139}
  {arXiv:hep-ph/0501139 [hep-ph]} \BibitemShut {NoStop}%
\bibitem [{\citenamefont {Ellwanger}\ \emph {et~al.}(2010)\citenamefont
  {Ellwanger}, \citenamefont {Hugonie},\ and\ \citenamefont
  {Teixeira}}]{Ellwanger:2009dp}%
  \BibitemOpen
  \bibfield  {author} {\bibinfo {author} {\bibfnamefont {U.}~\bibnamefont
  {Ellwanger}}, \bibinfo {author} {\bibfnamefont {C.}~\bibnamefont {Hugonie}},
  \ and\ \bibinfo {author} {\bibfnamefont {A.~M.}\ \bibnamefont {Teixeira}},\
  }\href {\doibase 10.1016/j.physrep.2010.07.001} {\bibfield  {journal}
  {\bibinfo  {journal} {Phys. Rept.}\ }\textbf {\bibinfo {volume} {496}},\
  \bibinfo {pages} {1} (\bibinfo {year} {2010})},\ \Eprint
  {http://arxiv.org/abs/0910.1785} {arXiv:0910.1785 [hep-ph]} \BibitemShut
  {NoStop}%
\bibitem [{\citenamefont {Maniatis}(2010)}]{Maniatis:2009re}%
  \BibitemOpen
  \bibfield  {author} {\bibinfo {author} {\bibfnamefont {M.}~\bibnamefont
  {Maniatis}},\ }\href {\doibase 10.1142/S0217751X10049827} {\bibfield
  {journal} {\bibinfo  {journal} {Int. J. Mod. Phys.}\ }\textbf {\bibinfo
  {volume} {A25}},\ \bibinfo {pages} {3505} (\bibinfo {year} {2010})},\ \Eprint
  {http://arxiv.org/abs/0906.0777} {arXiv:0906.0777 [hep-ph]} \BibitemShut
  {NoStop}%
\bibitem [{\citenamefont {Franke}\ and\ \citenamefont
  {Fraas}(1997)}]{Franke:1995tc}%
  \BibitemOpen
  \bibfield  {author} {\bibinfo {author} {\bibfnamefont {F.}~\bibnamefont
  {Franke}}\ and\ \bibinfo {author} {\bibfnamefont {H.}~\bibnamefont {Fraas}},\
  }\href {\doibase 10.1142/S0217751X97000529} {\bibfield  {journal} {\bibinfo
  {journal} {Int. J. Mod. Phys.}\ }\textbf {\bibinfo {volume} {A12}},\ \bibinfo
  {pages} {479} (\bibinfo {year} {1997})},\ \Eprint
  {http://arxiv.org/abs/hep-ph/9512366} {arXiv:hep-ph/9512366 [hep-ph]}
  \BibitemShut {NoStop}%
\bibitem [{\citenamefont {Ellwanger}(1993)}]{Ellwanger:1993hn}%
  \BibitemOpen
  \bibfield  {author} {\bibinfo {author} {\bibfnamefont {U.}~\bibnamefont
  {Ellwanger}},\ }\href {\doibase 10.1016/0370-2693(93)91431-L} {\bibfield
  {journal} {\bibinfo  {journal} {Phys. Lett.}\ }\textbf {\bibinfo {volume}
  {B303}},\ \bibinfo {pages} {271} (\bibinfo {year} {1993})},\ \Eprint
  {http://arxiv.org/abs/hep-ph/9302224} {arXiv:hep-ph/9302224 [hep-ph]}
  \BibitemShut {NoStop}%
\bibitem [{\citenamefont {Ellis}\ \emph {et~al.}(1989)\citenamefont {Ellis},
  \citenamefont {Gunion}, \citenamefont {Haber}, \citenamefont {Roszkowski},\
  and\ \citenamefont {Zwirner}}]{Ellis:1988er}%
  \BibitemOpen
  \bibfield  {author} {\bibinfo {author} {\bibfnamefont {J.~R.}\ \bibnamefont
  {Ellis}}, \bibinfo {author} {\bibfnamefont {J.~F.}\ \bibnamefont {Gunion}},
  \bibinfo {author} {\bibfnamefont {H.~E.}\ \bibnamefont {Haber}}, \bibinfo
  {author} {\bibfnamefont {L.}~\bibnamefont {Roszkowski}}, \ and\ \bibinfo
  {author} {\bibfnamefont {F.}~\bibnamefont {Zwirner}},\ }\href {\doibase
  10.1103/PhysRevD.39.844} {\bibfield  {journal} {\bibinfo  {journal} {Phys.
  Rev.}\ }\textbf {\bibinfo {volume} {D39}},\ \bibinfo {pages} {844} (\bibinfo
  {year} {1989})}\BibitemShut {NoStop}%
\bibitem [{\citenamefont {Derendinger}\ and\ \citenamefont
  {Savoy}(1984)}]{Derendinger:1983bz}%
  \BibitemOpen
  \bibfield  {author} {\bibinfo {author} {\bibfnamefont {J.~P.}\ \bibnamefont
  {Derendinger}}\ and\ \bibinfo {author} {\bibfnamefont {C.~A.}\ \bibnamefont
  {Savoy}},\ }\href {\doibase 10.1016/0550-3213(84)90162-7} {\bibfield
  {journal} {\bibinfo  {journal} {Nucl. Phys.}\ }\textbf {\bibinfo {volume}
  {B237}},\ \bibinfo {pages} {307} (\bibinfo {year} {1984})}\BibitemShut
  {NoStop}%
\bibitem [{\citenamefont {Nilles}\ \emph {et~al.}(1983)\citenamefont {Nilles},
  \citenamefont {Srednicki},\ and\ \citenamefont {Wyler}}]{Nilles:1982dy}%
  \BibitemOpen
  \bibfield  {author} {\bibinfo {author} {\bibfnamefont {H.~P.}\ \bibnamefont
  {Nilles}}, \bibinfo {author} {\bibfnamefont {M.}~\bibnamefont {Srednicki}}, \
  and\ \bibinfo {author} {\bibfnamefont {D.}~\bibnamefont {Wyler}},\ }\href
  {\doibase 10.1016/0370-2693(83)90460-4} {\bibfield  {journal} {\bibinfo
  {journal} {Phys. Lett.}\ }\textbf {\bibinfo {volume} {120B}},\ \bibinfo
  {pages} {346} (\bibinfo {year} {1983})}\BibitemShut {NoStop}%
\bibitem [{\citenamefont {Drees}(1989)}]{Drees:1988fc}%
  \BibitemOpen
  \bibfield  {author} {\bibinfo {author} {\bibfnamefont {M.}~\bibnamefont
  {Drees}},\ }\href {\doibase 10.1142/S0217751X89001448} {\bibfield  {journal}
  {\bibinfo  {journal} {Int. J. Mod. Phys.}\ }\textbf {\bibinfo {volume}
  {A4}},\ \bibinfo {pages} {3635} (\bibinfo {year} {1989})}\BibitemShut
  {NoStop}%
\bibitem [{\citenamefont {Khalil}\ and\ \citenamefont
  {Moretti}(2017)}]{Khalil:2132388}%
  \BibitemOpen
  \bibfield  {author} {\bibinfo {author} {\bibfnamefont {S.}~\bibnamefont
  {Khalil}}\ and\ \bibinfo {author} {\bibfnamefont {S.}~\bibnamefont
  {Moretti}},\ }\href {https://cds.cern.ch/record/2132388} {\emph {\bibinfo
  {title} {{Supersymmetry beyond minimality: from theory to experiment}}}}\
  (\bibinfo  {publisher} {CRC Press},\ \bibinfo {address} {Boca Raton, FL},\
  \bibinfo {year} {2017})\BibitemShut {NoStop}%
\bibitem [{\citenamefont {Aad}\ \emph {et~al.}(2015{\natexlab{a}})\citenamefont
  {Aad} \emph {et~al.}}]{Aad:2014kga}%
  \BibitemOpen
  \bibfield  {author} {\bibinfo {author} {\bibfnamefont {G.}~\bibnamefont
  {Aad}} \emph {et~al.} (\bibinfo {collaboration} {ATLAS}),\ }\href {\doibase
  10.1007/JHEP03(2015)088} {\bibfield  {journal} {\bibinfo  {journal} {JHEP}\
  }\textbf {\bibinfo {volume} {03}},\ \bibinfo {pages} {088} (\bibinfo {year}
  {2015}{\natexlab{a}})},\ \Eprint {http://arxiv.org/abs/1412.6663}
  {arXiv:1412.6663 [hep-ex]} \BibitemShut {NoStop}%
\bibitem [{\citenamefont {Khachatryan}\ \emph {et~al.}(2015)\citenamefont
  {Khachatryan} \emph {et~al.}}]{Khachatryan:2015qxa}%
  \BibitemOpen
  \bibfield  {author} {\bibinfo {author} {\bibfnamefont {V.}~\bibnamefont
  {Khachatryan}} \emph {et~al.} (\bibinfo {collaboration} {CMS}),\ }\href
  {\doibase 10.1007/JHEP11(2015)018} {\bibfield  {journal} {\bibinfo  {journal}
  {JHEP}\ }\textbf {\bibinfo {volume} {11}},\ \bibinfo {pages} {018} (\bibinfo
  {year} {2015})},\ \Eprint {http://arxiv.org/abs/1508.07774} {arXiv:1508.07774
  [hep-ex]} \BibitemShut {NoStop}%
\bibitem [{\citenamefont {Aad}\ \emph {et~al.}(2016{\natexlab{b}})\citenamefont
  {Aad} \emph {et~al.}}]{Aad:2015typ}%
  \BibitemOpen
  \bibfield  {author} {\bibinfo {author} {\bibfnamefont {G.}~\bibnamefont
  {Aad}} \emph {et~al.} (\bibinfo {collaboration} {ATLAS}),\ }\href {\doibase
  10.1007/JHEP03(2016)127} {\bibfield  {journal} {\bibinfo  {journal} {JHEP}\
  }\textbf {\bibinfo {volume} {03}},\ \bibinfo {pages} {127} (\bibinfo {year}
  {2016}{\natexlab{b}})},\ \Eprint {http://arxiv.org/abs/1512.03704}
  {arXiv:1512.03704 [hep-ex]} \BibitemShut {NoStop}%
\bibitem [{\citenamefont {Bergeaas~Kuutmann}(2017)}]{BergeaasKuutmann:2017yud}%
  \BibitemOpen
  \bibfield  {author} {\bibinfo {author} {\bibfnamefont {E.}~\bibnamefont
  {Bergeaas~Kuutmann}} (\bibinfo {collaboration} {ATLAS}),\ }\bibfield
  {booktitle} {\emph {\bibinfo {booktitle} {{Proceedings, 2017 European
  Physical Society Conference on High Energy Physics (EPS-HEP 2017): Venice,
  Italy, July 5-12, 2017}}},\ }\href {\doibase 10.22323/1.314.0260} {\bibfield
  {journal} {\bibinfo  {journal} {PoS}\ }\textbf {\bibinfo {volume}
  {EPS-HEP2017}},\ \bibinfo {pages} {260} (\bibinfo {year} {2017})}\BibitemShut
  {NoStop}%
\bibitem [{\citenamefont {Sirunyan}\ \emph {et~al.}(2017)\citenamefont
  {Sirunyan} \emph {et~al.}}]{Sirunyan:2017sbn}%
  \BibitemOpen
  \bibfield  {author} {\bibinfo {author} {\bibfnamefont {A.~M.}\ \bibnamefont
  {Sirunyan}} \emph {et~al.} (\bibinfo {collaboration} {CMS}),\ }\href
  {\doibase 10.1103/PhysRevLett.119.141802} {\bibfield  {journal} {\bibinfo
  {journal} {Phys. Rev. Lett.}\ }\textbf {\bibinfo {volume} {119}},\ \bibinfo
  {pages} {141802} (\bibinfo {year} {2017})},\ \Eprint
  {http://arxiv.org/abs/1705.02942} {arXiv:1705.02942 [hep-ex]} \BibitemShut
  {NoStop}%
\bibitem [{\citenamefont {Laurila}(2017)}]{Laurila:2017phk}%
  \BibitemOpen
  \bibfield  {author} {\bibinfo {author} {\bibfnamefont {S.}~\bibnamefont
  {Laurila}} (\bibinfo {collaboration} {CMS}),\ }\bibfield  {booktitle} {\emph
  {\bibinfo {booktitle} {{Proceedings, 6th International Workshop on Prospects
  for Charged Higgs Discovery at Colliders (CHARGED 2016): Uppsala, Sweden,
  October 3-6, 2016}}},\ }\href@noop {} {\bibfield  {journal} {\bibinfo
  {journal} {PoS}\ }\textbf {\bibinfo {volume} {CHARGED2016}},\ \bibinfo
  {pages} {008} (\bibinfo {year} {2017})}\BibitemShut {NoStop}%
\bibitem [{\citenamefont {Akeroyd}\ \emph {et~al.}(2017)\citenamefont {Akeroyd}
  \emph {et~al.}}]{Akeroyd:2016ymd}%
  \BibitemOpen
  \bibfield  {author} {\bibinfo {author} {\bibfnamefont {A.~G.}\ \bibnamefont
  {Akeroyd}} \emph {et~al.},\ }\href {\doibase 10.1140/epjc/s10052-017-4829-2}
  {\bibfield  {journal} {\bibinfo  {journal} {Eur. Phys. J.}\ }\textbf
  {\bibinfo {volume} {C77}},\ \bibinfo {pages} {276} (\bibinfo {year}
  {2017})},\ \Eprint {http://arxiv.org/abs/1607.01320} {arXiv:1607.01320
  [hep-ph]} \BibitemShut {NoStop}%
\bibitem [{Note1()}]{Note1}%
  \BibitemOpen
  \bibinfo {note} {In fact, also $h^\pm \to W^\pm Z$ decays have recently been
  searched for at the LHC \cite {Sirunyan:2017sbn}.}\BibitemShut {Stop}%
\bibitem [{\citenamefont {Diaz-Cruz}\ \emph {et~al.}(2009)\citenamefont
  {Diaz-Cruz}, \citenamefont {Hernandez-Sanchez}, \citenamefont {Moretti},
  \citenamefont {Noriega-Papaqui},\ and\ \citenamefont
  {Rosado}}]{DiazCruz:2009ek}%
  \BibitemOpen
  \bibfield  {author} {\bibinfo {author} {\bibfnamefont {J.~L.}\ \bibnamefont
  {Diaz-Cruz}}, \bibinfo {author} {\bibfnamefont {J.}~\bibnamefont
  {Hernandez-Sanchez}}, \bibinfo {author} {\bibfnamefont {S.}~\bibnamefont
  {Moretti}}, \bibinfo {author} {\bibfnamefont {R.}~\bibnamefont
  {Noriega-Papaqui}}, \ and\ \bibinfo {author} {\bibfnamefont {A.}~\bibnamefont
  {Rosado}},\ }\href {\doibase 10.1103/PhysRevD.79.095025} {\bibfield
  {journal} {\bibinfo  {journal} {Phys. Rev.}\ }\textbf {\bibinfo {volume}
  {D79}},\ \bibinfo {pages} {095025} (\bibinfo {year} {2009})},\ \Eprint
  {http://arxiv.org/abs/0902.4490} {arXiv:0902.4490 [hep-ph]} \BibitemShut
  {NoStop}%
\bibitem [{\citenamefont {Hernandez-Sanchez}\ \emph {et~al.}(2013)\citenamefont
  {Hernandez-Sanchez}, \citenamefont {Moretti}, \citenamefont
  {Noriega-Papaqui},\ and\ \citenamefont {Rosado}}]{HernandezSanchez:2012eg}%
  \BibitemOpen
  \bibfield  {author} {\bibinfo {author} {\bibfnamefont {J.}~\bibnamefont
  {Hernandez-Sanchez}}, \bibinfo {author} {\bibfnamefont {S.}~\bibnamefont
  {Moretti}}, \bibinfo {author} {\bibfnamefont {R.}~\bibnamefont
  {Noriega-Papaqui}}, \ and\ \bibinfo {author} {\bibfnamefont {A.}~\bibnamefont
  {Rosado}},\ }\href {\doibase 10.1007/JHEP07(2013)044} {\bibfield  {journal}
  {\bibinfo  {journal} {JHEP}\ }\textbf {\bibinfo {volume} {07}},\ \bibinfo
  {pages} {044} (\bibinfo {year} {2013})},\ \Eprint
  {http://arxiv.org/abs/1212.6818} {arXiv:1212.6818 [hep-ph]} \BibitemShut
  {NoStop}%
\bibitem [{\citenamefont {Akeroyd}\ \emph {et~al.}(2012)\citenamefont
  {Akeroyd}, \citenamefont {Moretti},\ and\ \citenamefont
  {Hernandez-Sanchez}}]{Akeroyd:2012yg}%
  \BibitemOpen
  \bibfield  {author} {\bibinfo {author} {\bibfnamefont {A.~G.}\ \bibnamefont
  {Akeroyd}}, \bibinfo {author} {\bibfnamefont {S.}~\bibnamefont {Moretti}}, \
  and\ \bibinfo {author} {\bibfnamefont {J.}~\bibnamefont
  {Hernandez-Sanchez}},\ }\href {\doibase 10.1103/PhysRevD.85.115002}
  {\bibfield  {journal} {\bibinfo  {journal} {Phys. Rev.}\ }\textbf {\bibinfo
  {volume} {D85}},\ \bibinfo {pages} {115002} (\bibinfo {year} {2012})},\
  \Eprint {http://arxiv.org/abs/1203.5769} {arXiv:1203.5769 [hep-ph]}
  \BibitemShut {NoStop}%
\bibitem [{\citenamefont {Collaboration}(2016)}]{CMS:2016qoa}%
  \BibitemOpen
  \bibfield  {author} {\bibinfo {author} {\bibfnamefont {C.}~\bibnamefont
  {Collaboration}} (\bibinfo {collaboration} {CMS}),\ }\href@noop {} {\bibfield
   {journal} {\bibinfo  {journal} {CMS-PAS-HIG-16-030}\ } (\bibinfo {year}
  {2016})}\BibitemShut {NoStop}%
\bibitem [{\citenamefont {Das}(2018)}]{Das:2018fog}%
  \BibitemOpen
  \bibfield  {author} {\bibinfo {author} {\bibfnamefont {D.}~\bibnamefont
  {Das}},\ }\href@noop {} {\  (\bibinfo {year} {2018})},\ \Eprint
  {http://arxiv.org/abs/1804.06630} {arXiv:1804.06630 [hep-ph]} \BibitemShut
  {NoStop}%
\bibitem [{\citenamefont {Abelleira~Fernandez}\ \emph
  {et~al.}(2012{\natexlab{a}})\citenamefont {Abelleira~Fernandez} \emph
  {et~al.}}]{AbelleiraFernandez:2012cc}%
  \BibitemOpen
  \bibfield  {author} {\bibinfo {author} {\bibfnamefont {J.~L.}\ \bibnamefont
  {Abelleira~Fernandez}} \emph {et~al.} (\bibinfo {collaboration} {LHeC Study
  Group}),\ }\href {\doibase 10.1088/0954-3899/39/7/075001} {\bibfield
  {journal} {\bibinfo  {journal} {J. Phys.}\ }\textbf {\bibinfo {volume}
  {G39}},\ \bibinfo {pages} {075001} (\bibinfo {year} {2012}{\natexlab{a}})},\
  \Eprint {http://arxiv.org/abs/1206.2913} {arXiv:1206.2913 [physics.acc-ph]}
  \BibitemShut {NoStop}%
\bibitem [{\citenamefont {Kuze}(2018)}]{Kuze:2018dqd}%
  \BibitemOpen
  \bibfield  {author} {\bibinfo {author} {\bibfnamefont {M.}~\bibnamefont
  {Kuze}},\ }in\ \href
  {https://inspirehep.net/record/1650011/files/arXiv:1801.07394.pdf} {\emph
  {\bibinfo {booktitle} {{21st International Conference on Particles and Nuclei
  (PANIC 17) Beijing, China, September 1-5, 2017}}}}\ (\bibinfo {year} {2018})\
  \Eprint {http://arxiv.org/abs/1801.07394} {arXiv:1801.07394 [hep-ex]}
  \BibitemShut {NoStop}%
\bibitem [{\citenamefont {Britzger}\ and\ \citenamefont
  {Klein}(2018)}]{Britzger:2017fuc}%
  \BibitemOpen
  \bibfield  {author} {\bibinfo {author} {\bibfnamefont {D.}~\bibnamefont
  {Britzger}}\ and\ \bibinfo {author} {\bibfnamefont {M.}~\bibnamefont
  {Klein}},\ }\bibfield  {booktitle} {\emph {\bibinfo {booktitle}
  {{Proceedings, 25th International Workshop on Deep-Inelastic Scattering and
  Related Topics (DIS 2017): Birmingham, UK, April 3-7, 2017}}},\ }\href@noop
  {} {\bibfield  {journal} {\bibinfo  {journal} {PoS}\ }\textbf {\bibinfo
  {volume} {DIS2017}},\ \bibinfo {pages} {105} (\bibinfo {year}
  {2018})}\BibitemShut {NoStop}%
\bibitem [{\citenamefont {Das}\ \emph {et~al.}(2016)\citenamefont {Das},
  \citenamefont {Hern{\'a}ndez-S{\'a}nchez}, \citenamefont {Moretti},
  \citenamefont {Rosado},\ and\ \citenamefont {Xoxocotzi}}]{Das:2015kea}%
  \BibitemOpen
  \bibfield  {author} {\bibinfo {author} {\bibfnamefont {S.~P.}\ \bibnamefont
  {Das}}, \bibinfo {author} {\bibfnamefont {J.}~\bibnamefont
  {Hern{\'a}ndez-S{\'a}nchez}}, \bibinfo {author} {\bibfnamefont
  {S.}~\bibnamefont {Moretti}}, \bibinfo {author} {\bibfnamefont
  {A.}~\bibnamefont {Rosado}}, \ and\ \bibinfo {author} {\bibfnamefont
  {R.}~\bibnamefont {Xoxocotzi}},\ }\href {\doibase 10.1103/PhysRevD.94.055003}
  {\bibfield  {journal} {\bibinfo  {journal} {Phys. Rev.}\ }\textbf {\bibinfo
  {volume} {D94}},\ \bibinfo {pages} {055003} (\bibinfo {year} {2016})},\
  \Eprint {http://arxiv.org/abs/1503.01464} {arXiv:1503.01464 [hep-ph]}
  \BibitemShut {NoStop}%
\bibitem [{\citenamefont {Hern{\'a}ndez-S{\'a}nchez}\ \emph
  {et~al.}(2017)\citenamefont {Hern{\'a}ndez-S{\'a}nchez}, \citenamefont
  {Flores-S{\'a}nchez}, \citenamefont {Honorato}, \citenamefont {Moretti},\
  and\ \citenamefont {Rosado}}]{Hernandez-Sanchez:2016vys}%
  \BibitemOpen
  \bibfield  {author} {\bibinfo {author} {\bibfnamefont {J.}~\bibnamefont
  {Hern{\'a}ndez-S{\'a}nchez}}, \bibinfo {author} {\bibfnamefont
  {O.}~\bibnamefont {Flores-S{\'a}nchez}}, \bibinfo {author} {\bibfnamefont
  {C.~G.}\ \bibnamefont {Honorato}}, \bibinfo {author} {\bibfnamefont
  {S.}~\bibnamefont {Moretti}}, \ and\ \bibinfo {author} {\bibfnamefont
  {S.}~\bibnamefont {Rosado}},\ }\bibfield  {booktitle} {\emph {\bibinfo
  {booktitle} {{Proceedings, 6th International Workshop on Prospects for
  Charged Higgs Discovery at Colliders (CHARGED 2016): Uppsala, Sweden, October
  3-6, 2016}}},\ }\href@noop {} {\bibfield  {journal} {\bibinfo  {journal}
  {PoS}\ }\textbf {\bibinfo {volume} {CHARGED2016}},\ \bibinfo {pages} {032}
  (\bibinfo {year} {2017})},\ \Eprint {http://arxiv.org/abs/1612.06316}
  {arXiv:1612.06316 [hep-ph]} \BibitemShut {NoStop}%
\bibitem [{\citenamefont {Mosomane}\ \emph {et~al.}(2017)\citenamefont
  {Mosomane}, \citenamefont {Kumar}, \citenamefont {Cornell},\ and\
  \citenamefont {Mellado}}]{Mosomane:2017jcg}%
  \BibitemOpen
  \bibfield  {author} {\bibinfo {author} {\bibfnamefont {C.}~\bibnamefont
  {Mosomane}}, \bibinfo {author} {\bibfnamefont {M.}~\bibnamefont {Kumar}},
  \bibinfo {author} {\bibfnamefont {A.~S.}\ \bibnamefont {Cornell}}, \ and\
  \bibinfo {author} {\bibfnamefont {B.}~\bibnamefont {Mellado}},\ }\bibfield
  {booktitle} {\emph {\bibinfo {booktitle} {{Proceedings, Workshop on High
  Energy Particle Physics (HEPPW2017): Johannesburg, South Africa, February
  1-3, 2017}}},\ }\href {\doibase 10.1088/1742-6596/889/1/012004} {\bibfield
  {journal} {\bibinfo  {journal} {J. Phys. Conf. Ser.}\ }\textbf {\bibinfo
  {volume} {889}},\ \bibinfo {pages} {012004} (\bibinfo {year} {2017})},\
  \Eprint {http://arxiv.org/abs/1707.05997} {arXiv:1707.05997 [hep-ph]}
  \BibitemShut {NoStop}%
\bibitem [{\citenamefont {Das}\ \emph {et~al.}(2017)\citenamefont {Das},
  \citenamefont {Fraga},\ and\ \citenamefont {Avila}}]{Das:2017mqw}%
  \BibitemOpen
  \bibfield  {author} {\bibinfo {author} {\bibfnamefont {S.~P.}\ \bibnamefont
  {Das}}, \bibinfo {author} {\bibfnamefont {J.}~\bibnamefont {Fraga}}, \ and\
  \bibinfo {author} {\bibfnamefont {C.}~\bibnamefont {Avila}},\ }\href@noop {}
  {\  (\bibinfo {year} {2017})},\ \Eprint {http://arxiv.org/abs/1712.04395}
  {arXiv:1712.04395 [hep-ph]} \BibitemShut {NoStop}%
\bibitem [{\citenamefont {Zheng}(2015)}]{Zheng:2014loa}%
  \BibitemOpen
  \bibfield  {author} {\bibinfo {author} {\bibfnamefont {S.}~\bibnamefont
  {Zheng}},\ }\href {\doibase 10.1140/epjc/s10052-015-3416-7} {\bibfield
  {journal} {\bibinfo  {journal} {Eur. Phys. J.}\ }\textbf {\bibinfo {volume}
  {C75}},\ \bibinfo {pages} {195} (\bibinfo {year} {2015})},\ \Eprint
  {http://arxiv.org/abs/1405.6907} {arXiv:1405.6907 [hep-ph]} \BibitemShut
  {NoStop}%
\bibitem [{\citenamefont {Ade}\ \emph {et~al.}(2016)\citenamefont {Ade} \emph
  {et~al.}}]{Ade:2015xua}%
  \BibitemOpen
  \bibfield  {author} {\bibinfo {author} {\bibfnamefont {P.~A.~R.}\
  \bibnamefont {Ade}} \emph {et~al.} (\bibinfo {collaboration} {Planck}),\
  }\href {\doibase 10.1051/0004-6361/201525830} {\bibfield  {journal} {\bibinfo
   {journal} {Astron. Astrophys.}\ }\textbf {\bibinfo {volume} {594}},\
  \bibinfo {pages} {A13} (\bibinfo {year} {2016})},\ \Eprint
  {http://arxiv.org/abs/1502.01589} {arXiv:1502.01589 [astro-ph.CO]}
  \BibitemShut {NoStop}%
\bibitem [{\citenamefont {Drees}\ and\ \citenamefont
  {Gerbier}(2012)}]{Drees:2012ji}%
  \BibitemOpen
  \bibfield  {author} {\bibinfo {author} {\bibfnamefont {M.}~\bibnamefont
  {Drees}}\ and\ \bibinfo {author} {\bibfnamefont {G.}~\bibnamefont
  {Gerbier}},\ }\href@noop {} {\  (\bibinfo {year} {2012})},\ \Eprint
  {http://arxiv.org/abs/1204.2373} {arXiv:1204.2373 [hep-ph]} \BibitemShut
  {NoStop}%
\bibitem [{\citenamefont {Sanabria}(2014)}]{Sanabria:2014yva}%
  \BibitemOpen
  \bibfield  {author} {\bibinfo {author} {\bibfnamefont {J.~C.}\ \bibnamefont
  {Sanabria}},\ }\href@noop {} {\bibfield  {journal} {\bibinfo  {journal} {Rev.
  Acad. Colomb. Cienc.}\ }\textbf {\bibinfo {volume} {38}},\ \bibinfo {pages}
  {34} (\bibinfo {year} {2014})}\BibitemShut {NoStop}%
\bibitem [{\citenamefont {Khachatryan}\ \emph {et~al.}(2017)\citenamefont
  {Khachatryan} \emph {et~al.}}]{Khachatryan:2016whc}%
  \BibitemOpen
  \bibfield  {author} {\bibinfo {author} {\bibfnamefont {V.}~\bibnamefont
  {Khachatryan}} \emph {et~al.} (\bibinfo {collaboration} {CMS}),\ }\href
  {\doibase 10.1007/JHEP02(2017)135} {\bibfield  {journal} {\bibinfo  {journal}
  {JHEP}\ }\textbf {\bibinfo {volume} {02}},\ \bibinfo {pages} {135} (\bibinfo
  {year} {2017})},\ \Eprint {http://arxiv.org/abs/1610.09218} {arXiv:1610.09218
  [hep-ex]} \BibitemShut {NoStop}%
\bibitem [{\citenamefont {Aad}\ \emph {et~al.}(2015{\natexlab{b}})\citenamefont
  {Aad} \emph {et~al.}}]{Aad:2015pla}%
  \BibitemOpen
  \bibfield  {author} {\bibinfo {author} {\bibfnamefont {G.}~\bibnamefont
  {Aad}} \emph {et~al.} (\bibinfo {collaboration} {ATLAS}),\ }\href {\doibase
  10.1007/JHEP11(2015)206} {\bibfield  {journal} {\bibinfo  {journal} {JHEP}\
  }\textbf {\bibinfo {volume} {11}},\ \bibinfo {pages} {206} (\bibinfo {year}
  {2015}{\natexlab{b}})},\ \Eprint {http://arxiv.org/abs/1509.00672}
  {arXiv:1509.00672 [hep-ex]} \BibitemShut {NoStop}%
\bibitem [{\citenamefont {Das}\ \emph {et~al.}()\citenamefont {Das},
  \citenamefont {Hern\'andez-S\'anchez}, \citenamefont {Moretti},\ and\
  \citenamefont {Rosado}}]{rhmd}%
  \BibitemOpen
  \bibfield  {author} {\bibinfo {author} {\bibfnamefont {S.~P.}\ \bibnamefont
  {Das}}, \bibinfo {author} {\bibfnamefont {J.}~\bibnamefont
  {Hern\'andez-S\'anchez}}, \bibinfo {author} {\bibfnamefont {S.}~\bibnamefont
  {Moretti}}, \ and\ \bibinfo {author} {\bibfnamefont {A.}~\bibnamefont
  {Rosado}},\ }\href@noop {} {\ }\Eprint {http://arxiv.org/abs/in preparation}
  {in preparation} \BibitemShut {NoStop}%
\bibitem [{\citenamefont {Alwall}\ \emph {et~al.}(2014)\citenamefont {Alwall},
  \citenamefont {Frederix}, \citenamefont {Frixione}, \citenamefont {Hirschi},
  \citenamefont {Maltoni}, \citenamefont {Mattelaer}, \citenamefont {Shao},
  \citenamefont {Stelzer}, \citenamefont {Torrielli},\ and\ \citenamefont
  {Zaro}}]{Alwall:2014hca}%
  \BibitemOpen
  \bibfield  {author} {\bibinfo {author} {\bibfnamefont {J.}~\bibnamefont
  {Alwall}}, \bibinfo {author} {\bibfnamefont {R.}~\bibnamefont {Frederix}},
  \bibinfo {author} {\bibfnamefont {S.}~\bibnamefont {Frixione}}, \bibinfo
  {author} {\bibfnamefont {V.}~\bibnamefont {Hirschi}}, \bibinfo {author}
  {\bibfnamefont {F.}~\bibnamefont {Maltoni}}, \bibinfo {author} {\bibfnamefont
  {O.}~\bibnamefont {Mattelaer}}, \bibinfo {author} {\bibfnamefont {H.~S.}\
  \bibnamefont {Shao}}, \bibinfo {author} {\bibfnamefont {T.}~\bibnamefont
  {Stelzer}}, \bibinfo {author} {\bibfnamefont {P.}~\bibnamefont {Torrielli}},
  \ and\ \bibinfo {author} {\bibfnamefont {M.}~\bibnamefont {Zaro}},\ }\href
  {\doibase 10.1007/JHEP07(2014)079} {\bibfield  {journal} {\bibinfo  {journal}
  {JHEP}\ }\textbf {\bibinfo {volume} {07}},\ \bibinfo {pages} {079} (\bibinfo
  {year} {2014})},\ \Eprint {http://arxiv.org/abs/1405.0301} {arXiv:1405.0301
  [hep-ph]} \BibitemShut {NoStop}%
\bibitem [{\citenamefont {Bruening}\ and\ \citenamefont
  {Klein}(2013)}]{Bruening:2013bga}%
  \BibitemOpen
  \bibfield  {author} {\bibinfo {author} {\bibfnamefont {O.}~\bibnamefont
  {Bruening}}\ and\ \bibinfo {author} {\bibfnamefont {M.}~\bibnamefont
  {Klein}},\ }\href {\doibase 10.1142/S0217732313300115} {\bibfield  {journal}
  {\bibinfo  {journal} {Mod. Phys. Lett.}\ }\textbf {\bibinfo {volume} {A28}},\
  \bibinfo {pages} {1330011} (\bibinfo {year} {2013})},\ \Eprint
  {http://arxiv.org/abs/1305.2090} {arXiv:1305.2090 [physics.acc-ph]}
  \BibitemShut {NoStop}%
\bibitem [{\citenamefont {Abelleira~Fernandez}\ \emph
  {et~al.}(2012{\natexlab{b}})\citenamefont {Abelleira~Fernandez} \emph
  {et~al.}}]{AbelleiraFernandez:2012ty}%
  \BibitemOpen
  \bibfield  {author} {\bibinfo {author} {\bibfnamefont {J.~L.}\ \bibnamefont
  {Abelleira~Fernandez}} \emph {et~al.} (\bibinfo {collaboration} {LHeC Study
  Group}),\ }\href@noop {} {\  (\bibinfo {year} {2012}{\natexlab{b}})},\
  \Eprint {http://arxiv.org/abs/1211.5102} {arXiv:1211.5102 [hep-ex]}
  \BibitemShut {NoStop}%
\bibitem [{\citenamefont {Appleby}\ \emph {et~al.}(2013)\citenamefont
  {Appleby}, \citenamefont {Thompson}, \citenamefont {Holzer}, \citenamefont
  {Fitterer}, \citenamefont {Bernard},\ and\ \citenamefont
  {Kostka}}]{Appleby:2013sha}%
  \BibitemOpen
  \bibfield  {author} {\bibinfo {author} {\bibfnamefont {R.~B.}\ \bibnamefont
  {Appleby}}, \bibinfo {author} {\bibfnamefont {L.}~\bibnamefont {Thompson}},
  \bibinfo {author} {\bibfnamefont {B.}~\bibnamefont {Holzer}}, \bibinfo
  {author} {\bibfnamefont {M.}~\bibnamefont {Fitterer}}, \bibinfo {author}
  {\bibfnamefont {N.}~\bibnamefont {Bernard}}, \ and\ \bibinfo {author}
  {\bibfnamefont {P.}~\bibnamefont {Kostka}},\ }\href {\doibase
  10.1088/0954-3899/40/12/125004} {\bibfield  {journal} {\bibinfo  {journal}
  {J. Phys.}\ }\textbf {\bibinfo {volume} {G40}},\ \bibinfo {pages} {125004}
  (\bibinfo {year} {2013})}\BibitemShut {NoStop}%
\bibitem [{\citenamefont {Ball}\ \emph {et~al.}(2013)\citenamefont {Ball},
  \citenamefont {Bertone}, \citenamefont {Carrazza}, \citenamefont
  {Del~Debbio}, \citenamefont {Forte}, \citenamefont {Guffanti}, \citenamefont
  {Hartland},\ and\ \citenamefont {Rojo}}]{Ball:2013hta}%
  \BibitemOpen
  \bibfield  {author} {\bibinfo {author} {\bibfnamefont {R.~D.}\ \bibnamefont
  {Ball}}, \bibinfo {author} {\bibfnamefont {V.}~\bibnamefont {Bertone}},
  \bibinfo {author} {\bibfnamefont {S.}~\bibnamefont {Carrazza}}, \bibinfo
  {author} {\bibfnamefont {L.}~\bibnamefont {Del~Debbio}}, \bibinfo {author}
  {\bibfnamefont {S.}~\bibnamefont {Forte}}, \bibinfo {author} {\bibfnamefont
  {A.}~\bibnamefont {Guffanti}}, \bibinfo {author} {\bibfnamefont {N.~P.}\
  \bibnamefont {Hartland}}, \ and\ \bibinfo {author} {\bibfnamefont
  {J.}~\bibnamefont {Rojo}} (\bibinfo {collaboration} {NNPDF}),\ }\href
  {\doibase 10.1016/j.nuclphysb.2013.10.010} {\bibfield  {journal} {\bibinfo
  {journal} {Nucl. Phys.}\ }\textbf {\bibinfo {volume} {B877}},\ \bibinfo
  {pages} {290} (\bibinfo {year} {2013})},\ \Eprint
  {http://arxiv.org/abs/1308.0598} {arXiv:1308.0598 [hep-ph]} \BibitemShut
  {NoStop}%
\bibitem [{\citenamefont {Pumplin}\ \emph {et~al.}(2002)\citenamefont
  {Pumplin}, \citenamefont {Stump}, \citenamefont {Huston}, \citenamefont
  {Lai}, \citenamefont {Nadolsky},\ and\ \citenamefont
  {Tung}}]{Pumplin:2002vw}%
  \BibitemOpen
  \bibfield  {author} {\bibinfo {author} {\bibfnamefont {J.}~\bibnamefont
  {Pumplin}}, \bibinfo {author} {\bibfnamefont {D.~R.}\ \bibnamefont {Stump}},
  \bibinfo {author} {\bibfnamefont {J.}~\bibnamefont {Huston}}, \bibinfo
  {author} {\bibfnamefont {H.~L.}\ \bibnamefont {Lai}}, \bibinfo {author}
  {\bibfnamefont {P.~M.}\ \bibnamefont {Nadolsky}}, \ and\ \bibinfo {author}
  {\bibfnamefont {W.~K.}\ \bibnamefont {Tung}},\ }\href {\doibase
  10.1088/1126-6708/2002/07/012} {\bibfield  {journal} {\bibinfo  {journal}
  {JHEP}\ }\textbf {\bibinfo {volume} {07}},\ \bibinfo {pages} {012} (\bibinfo
  {year} {2002})},\ \Eprint {http://arxiv.org/abs/hep-ph/0201195}
  {arXiv:hep-ph/0201195 [hep-ph]} \BibitemShut {NoStop}%
\bibitem [{\citenamefont {Sjostrand}\ \emph {et~al.}(2006)\citenamefont
  {Sjostrand}, \citenamefont {Mrenna},\ and\ \citenamefont
  {Skands}}]{Sjostrand:2006za}%
  \BibitemOpen
  \bibfield  {author} {\bibinfo {author} {\bibfnamefont {T.}~\bibnamefont
  {Sjostrand}}, \bibinfo {author} {\bibfnamefont {S.}~\bibnamefont {Mrenna}}, \
  and\ \bibinfo {author} {\bibfnamefont {P.~Z.}\ \bibnamefont {Skands}},\
  }\href {\doibase 10.1088/1126-6708/2006/05/026} {\bibfield  {journal}
  {\bibinfo  {journal} {JHEP}\ }\textbf {\bibinfo {volume} {05}},\ \bibinfo
  {pages} {026} (\bibinfo {year} {2006})},\ \Eprint
  {http://arxiv.org/abs/hep-ph/0603175} {arXiv:hep-ph/0603175 [hep-ph]}
  \BibitemShut {NoStop}%
\bibitem [{\citenamefont {Das}\ and\ \citenamefont
  {Nowakowski}(2017)}]{Das:2016eob}%
  \BibitemOpen
  \bibfield  {author} {\bibinfo {author} {\bibfnamefont {S.~P.}\ \bibnamefont
  {Das}}\ and\ \bibinfo {author} {\bibfnamefont {M.}~\bibnamefont
  {Nowakowski}},\ }\href {\doibase 10.1103/PhysRevD.96.055014} {\bibfield
  {journal} {\bibinfo  {journal} {Phys. Rev.}\ }\textbf {\bibinfo {volume}
  {D96}},\ \bibinfo {pages} {055014} (\bibinfo {year} {2017})},\ \Eprint
  {http://arxiv.org/abs/1612.07241} {arXiv:1612.07241 [hep-ph]} \BibitemShut
  {NoStop}%
\bibitem [{Note2()}]{Note2}%
  \BibitemOpen
  \bibinfo {note} {The efficiencies quoted in these numerical sections are
  always with respect to the previous set of selections.}\BibitemShut {Stop}%
\end{thebibliography}%
\end{document}